\documentclass[
  journal=largetwo,
  manuscript=article-type,
  year=2020,
  volume=37,
]{cup-journal}

\usepackage{aas-macros}
\usepackage{amsmath, amssymb}
\usepackage[nopatch]{microtype}
\usepackage{booktabs}

\usepackage{multirow}
\usepackage{graphicx, float}	
\usepackage{subcaption} 
\usepackage{parcolumns} 

\usepackage{hyperref}
\hypersetup{
  colorlinks   = true, 
  urlcolor     = blue,
  linkcolor    = blue, 
  citecolor    = blue
}

\newcommand{\DMISM}{DM$_{\rm ISM}$}

\newcommand{\DMEG}{DM$_{\rm EG}$}
\newcommand{\DMG}{DM$_{\rm MW}$}
\newcommand{\DMhalo}{DM$_{\rm halo}$}
\newcommand{\DMobs}{DM$_{\rm obs}$}

\newcommand{\DMunit}{pc\:cm$^{-3}$}

\newcommand{\nsfr}{$n_{\mathrm{sfr}}$}
\newcommand{\Emax}{$E_{\mathrm{max}}$}
\newcommand{\Emin}{$E_{\mathrm{min}}$}

\newcommand{\FAST}{FAST}

\newcommand{\flyseye}{CRAFT Fly's Eye}

\newcommand{\ics}{CRAFT ICS}
\newcommand{\parkes}{Murriyang (Parkes)}
\newcommand{\CHIMEa}{CHIME dec bin 1}
\newcommand{\CHIMEb}{CHIME dec bin 2}
\newcommand{\CHIMEc}{CHIME dec bin 3}
\newcommand{\CHIMEd}{CHIME dec bin 4}
\newcommand{\CHIMEe}{CHIME dec bin 5}
\newcommand{\CHIMEf}{CHIME dec bin 6}

\newcommand{\tfield}{\ensuremath{T_{\rm field}}}

\newcommand{\teff}{\ensuremath{T_{\rm eff}}}
\newcommand{\neff}{\ensuremath{N_{\rm eff}}}

\newcommand{\Rmin}{$R_{\mathrm{min}}$}
\newcommand{\Rmax}{$R_{\mathrm{max}}$}
\newcommand{\Rgamma}{$R_{\gamma}$}

\newcommand{\Rgammaval}{$-2.20^{+0.12}_{-0.22}$}
\newcommand{\Rminval}{$-4.23^{+1.10}_{-1.74}$\,d$^{-1}$}
\newcommand{\Rmaxval}{$1$\,d$^{-1}$}

\newcommand{\zdm}{\texttt{zDM}}

\title{A multi-telescope fit to the FRB population, allowing for FRB repetition}

\author{J.~Hoffmann}
\affiliation{International Centre for Radio Astronomy Research, Curtin University, Bentley, WA 6102, Australia}
\email[J. Hoffmann]{jordan.hoffmann@postgrad.curtin.edu.au}

\author{C.~W.~James}
\affiliation{International Centre for Radio Astronomy Research, Curtin University, Bentley, WA 6102, Australia}

\author{J.~X.~Prochaska}
\affiliation{Department of Astronomy and Astrophysics, University of California, Santa Cruz, CA 95064, USA}
\alsoaffiliation{Kavli Institute for the Physics and Mathematics of the Universe, 5-1-5 Kashiwanoha, Kashiwa 277-8583, Japan}
\alsoaffiliation{Division of Science, National Astronomical Observatory of Japan, 2-21-1 Osawa, Mitaka, Tokyo 181-8588, Japan}

\author{I.~Pastor-Marazuela}
\affiliation{ASTRON, Netherlands Institute for Radio Astronomy, Oude Hoogeveensedijk 4, 7991 PD Dwingeloo, The Netherlands}
\alsoaffiliation{Anton Pannekoek Institute for Astronomy, University of Amsterdam, Science Park 904, 1098 XH, Amsterdam, The Netherlands}
\alsoaffiliation{Jodrell Bank Centre for Astrophysics, Department of Physics and Astronomy, University of Manchester, Oxford Road, Manchester M13 9PL, UK}

\author{K.~Sharma}
\affiliation{Cahill Center for Astronomy and Astrophysics, MC 249-17 California Institute of Technology, Pasadena CA 91125, USA.}

\author{L.~Connor}
\affiliation{Center for Astrophysics $|$ Harvard \& Smithsonian, 60 Garden Street, Cambridge, MA 02138, USA}


\begin{document}

\begin{abstract}
Fast radio bursts (FRBs) are millisecond-duration radio pulses with extragalactic origins. A majority of FRBs have only had a single pulse detected from them; however, an increasing number have been identified as repeaters. Here we present a full redshift -- dispersion measure ($z$-DM) analysis including a simple power-law model of repetition rates. We assume all FRBs are drawn from the same intrinsic population and fit these FRB population parameters using data from ASKAP, Murriyang (Parkes), DSA, FAST, MeerKAT and CHIME catalogue 1. We do not utilise data from CHIME catalogue 2 yet, as we cannot robustly identify all repeaters in the sample. We constrain the minimum repetition rate above $10^{39}$\,erg to \Rmin{}$=$\Rminval{}, the maximum repetition rate to \Rmax{}$\gtrsim$\Rmaxval{} and the repetition rate index to \Rgamma{}$=$\Rgammaval{}. These parameters do not show significant correlation with any other FRB parameters. Our values are an improvement on existing results and are self-consistent with the DM distribution of CHIME and observed ratios of repeaters to non-repeaters across the sky. We also provide updated estimates of the FRB energy function and host galaxy parameters.
\end{abstract}

\section{Introduction} \label{sec:Introduction}
Fast radio bursts (FRBs) are highly energetic, extragalactic events that have been observed with a vast array of differing phenomenology \citep[e.g.][]{Sherman2023, CHIMEbaseband2024, Scott2025HTR}. Due to their cosmological origin, it is not possible to directly observe the progenitors of these bursts, and hence we must rely on observational signatures of the bursts themselves to deduce the production mechanism and thereby the progenitor object. However, it is difficult to reconcile a single progenitor model due to the incredible variety of burst morphologies and characteristics. As such, a plethora of FRB models have been suggested, with the most popular being a magnetar origin, but the exact progenitor still remains an open question \citep[e.g.][]{Zhang2023}.

Of the varying observational properties of FRBs, one of the most distinguishing factors is the apparent existence of two classes of FRBs: repeaters, in which multiple bursts are detected from a single object \citep[e.g.][]{Spitler2016, CHIME2023Cat1Reps}, and non-repeaters, in which only a single burst has been detected to date. Observationally, bursts from repeaters typically have a narrower bandwidth and more intricate temporal structure than non-repeaters \citep{Pleunis2021morphology}, although such a statement is not universal. Conversely, population synthesis studies have found that data from the Canadian Hydrogen Intensity (CHIME) can be modelled as a single population in which all FRBs are repeaters \citep{James2023, Cook2026_cat2repeaters}. However, neither result can conclude with certainty whether these do indeed form two distinct classes of FRBs, or if all FRBs are in fact repeaters and some have not been observed long enough to detect a second burst.

Population modelling is one of the most powerful methods to find overarching trends in the FRB population, thereby enabling us to link observational data to real physical models. Thus, this pathway provides the potential to help understand the mysterious origins of these bursts. In particular, FRB population studies help constrain the energetics, evolution and host-environment properties of FRBs, while also allowing us to study the gas through which the signal propagates. This second point has allowed FRB research to find incredible success in probing the gas of our Universe and has aided in cosmological studies \citep[e.g.][]{Macquart2020, Baptista2023, Khrykin2024, Connor2025, Sharma2026}. One of the greatest advantages of this method is that it allows several unknown parameters to be fit simultaneously, thereby eliminating the need for additional assumptions.

Many FRB population studies have been conducted to date \citep[e.g.][]{Gardenier2019_FRBpoppy, james2022, shin2023}, but they are broadly limited by assumptions in the models used and statistical noise due to a low number of localised FRBs with a determined redshift ($z$). While the issue of statistical noise will naturally be solved as we continually improve our detection systems and have more time to detect these objects, our model choices and the systematics contained within them will become an increasingly important issue.

As highlighted by \citet{James2023}, when modelling the population of FRBs, ignoring repetition would lead to an observational bias in our models. Specifically, repeaters are much easier to identify at low redshifts, as there are a larger number of visible bursts and hence the probability of detecting a second burst is higher. However, at high redshifts, only the brightest bursts are detectable and hence all FRBs tend towards being apparent non-repeaters, changing the distribution of FRBs between low and high redshifts.

In this work, we conduct a joint population study of a number of surveys while simultaneously including repetition statistics. This is the first time that repetition statistics have also been included in such a population study. In Section \ref{sec:method} we explain our methodology. Section \ref{sec:data} describes the data that we use and the models for each newly included telescope configuration and survey. Section \ref{sec:results} discusses our results and constraints on the repetition parameters, and lastly, we conclude in Section \ref{sec:conclusion}.


\section{Methodology} \label{sec:method}
The aim of this work is to incorporate repetition modelling into a full \zdm{} analysis. We use the same repetition model as \citet{James2023} and simultaneously model data from a number of different telescopes described in Section \ref{sec:data}.

The repetition model assumes all FRBs are derived from a single population and hence can be described by a single set of population parameters. This implicitly assumes all FRBs are intrinsic repeaters. In addition, we model each repeater as having some repetition rate $R$ above $10^{39}$\,ergs over a 1\,GHz band, or above $10^{30}$\,ergs\,Hz$^{-1}$ in the host frame. This is drawn from a power-law distribution with maximum rate \Rmax{}, minimum rate \Rmin{}, and slope \Rgamma{}. All three of these repetition parameters become free parameters that we fit in our analysis. To constrain these parameters, we compute a $P(z, \mathrm{DM})$ grid for single bursts and repeaters separately (but using the same population parameters). We also include energetics of the bursts by using the signal-to-noise ratio (SNR) at detection to compute $P(z,\mathrm{DM}, \mathrm{SNR})$ for each FRB using the respective grid with the methods described in \citet{james2022} and \citet{Hoffmann2025}. 

\subsection{Constraining repetition parameters} \label{sec:rep_params}
In order to understand the repetition statistics of a survey, it is necessary to know the time spent observing each field, $T_{\mathrm{field}}$, as spending more time on a given field significantly increases the probability of a second burst being detected from any repeater in that field. This therefore changes the probability of FRBs being classified as repeaters and places more stringent constraints on the repetition rate of each FRB in the field. For each survey where we know $T_{\mathrm{field}}$, we therefore construct two $P(z, \mathrm{DM})$ `grids' (one for single bursts and one for repeaters), and for each survey in which we do not know $T_{\mathrm{field}}$, a single $P(z, \mathrm{DM})$ grid is computed. For each such grid, we then calculate the total number of expected FRBs, $N_{\mathrm{exp}}$, by integrating each grid and fitting for a single normalisation constant that best matches all surveys simultaneously. We then compare this to the total number of observed FRBs ($N_{\mathrm{obs}} = N_{\mathrm{repeaters}} + N_{\mathrm{non-repeaters}}$) and, by assuming Poissonian statistics, we determine $P(N)$. The relative fraction of repeaters and single bursts is therefore incorporated into this $P(N)$ calculation, which is one of the primary constraining factors for the repetition parameters.

We additionally calculate $P(N_{\mathrm{b,obs}})$ for each repeater, where $N_{\mathrm{b,obs}}$ is the number of bursts detected from a specific repeater. Again, this is modelled as a Poisson process, which is an oversimplification.

\subsection{Including exact beam values}
For most surveys, we do not know the exact position that each FRB was detected within the beam. As such, we integrate over all possible beam values when determining the probability of the FRB being detected at the given SNR, $P({\mathrm{SNR}})$. For most surveys we consider, the beam values are fairly consistent over a large portion of the field of view and hence, while this does discard some information, it would not have a large impact on our results. The most significantly impacted parameters would be those that rely heavily on extreme events, such as the minimum and maximum energies (\Emin{} and \Emax{}) used as parameters in our luminosity distribution.

We have data on the exact position in the beam that FRBs were detected by the Australian SKA Pathfinder \citep[ASKAP;][]{deboer2009} (ASKAP) telescope during the Commensal Real-time ASKAP Fast Transients \citep[CRAFT;][]{Macquart2010} survey in incoherent sum mode (\ics{}). Hence, in this analysis, we simply use these values rather than integrating over the entire beam. For the other surveys, these values are not available, and hence we continue to average over all possible values as we previously have been.

\section{Survey description and data} \label{sec:data}
In this work, we use the same base data set as in \citet{Hoffmann2026} which includes data collected with ASKAP as part of the CRAFT survey in both the Fly's Eye (\flyseye{}) and \ics{} modes, the Murriyang (Parkes) telescope in its multibeam configuration \citep{StaveleySmith1996}, the Five-hundred-meter Aperture Spherical radio Telescope \citep[\FAST{};][]{FAST, FAST} and the Deep Synoptic Array \citep[DSA;][]{Ravi2021}. 

We include additional data from DSA, and also model surveys conducted with the MeerKAT \citep[More (meer) Karoo Array Telescope;][]{jonas2016} radio telescope and the Canadian Hydrogen Intensity Mapping Experiment \citep[CHIME;][]{CHIME}, which we did not previously incorporate. We use a modified data set of the \ics{} survey to allow us to account for repetition statistics within this survey. A more detailed description of the exact data sets we use is given later in this section. 

\subsection{Sample cuts}
When using redshift information, we can incur biases due to some sub-regions of the parameter space being easier to obtain redshifts for. In particular, we consider two cases in which using redshift information would create a bias in our analysis. Firstly, FRBs which have a higher redshift will on average have a fainter host galaxy and thereby be more difficult to detect. As such, if we have a sample of FRBs which have had optical follow-up, it is likely that the high-$z$ FRBs will not have redshifts. This changes the distribution of $P(z|\mathrm{DM_{EG}})$ for each extragalactic DM (\DMEG{}), hence creating a bias. If there is some \DMEG{} such that all FRBs above this value do not have redshifts and all FRBs below this value do have redshifts, then we can avoid this bias. As such, it is necessary to place cuts on the utilised redshifts such that we ignore redshifts where the extragalactic DM ($\rm DM_{EG}$) exceeds that of the lowest, unlocalised FRB \citep[see][for more details]{Hoffmann2026}. We additionally note that no FRBs have guaranteed host associations, and so here we implicitly make the assumption that all associations with probability of association $P(O) > 90\%$ actually have $P(O) = 100\%$ and all associations with $P(O) < 90\%$ have $P(O)=0\%$. We are now able to include the exact host association probabilities in our analyses, and so it is possible to more accurately model this in the future, although it is unlikely we will incorporate host redshifts with $P(O) < 10\%$ \citep{James2026_path_zdm}.


Similarly, it is also easier to obtain redshifts for stronger repeaters where localisation accuracy is a limiting factor in the original detections. This case is significant for the CHIME sample, as in general, the localisation accuracy is not sufficient to confidently associate FRBs with a unique host, and hence it is necessary to detect repeat bursts in follow-up observations to obtain a confident host galaxy. As such, it is necessary to select a complete region of the parameter space for each survey. This generally means we only use redshifts for repeaters exceeding some minimum number of repetitions, $N_{\mathrm{b,min}}$, where all FRBs with more bursts than $N_{\mathrm{b,min}}$ have associated redshifts.

Similarly to \citet{Hoffmann2026}, we place a latitude cut to exclude FRBs with $|b| < 20^\circ$ due to the large uncertainties in Galactic DM contributions (\DMG{}) through the Galactic plane. For DSA, we increase this constraint to $|b| < 34^\circ$ as this allows us to substantially increase the number of redshifts used (see Section \ref{sec:dsa} for more information).

The overall data that we use is summarised in Tables \ref{table:rep_surveys} and \ref{table:surveys}, which detail the surveys in which repetition statistics are accounted for and those in which we do not consider repetition at all. This division is necessary as the probability of detecting a repeater is highly dependent on how long a particular field is observed. The longer a field is observed, the higher the probability of detecting a second burst and therefore of classifying the FRB as a repeater. As such, surveys for which we do not have good estimates of the dwell time within each field cannot be used for repetition statistics.

\begin{table}
\begin{center}
\caption{The total number of FRBs and the corresponding number of redshifts and repeaters that we use in our analysis from each survey after the given cuts. We consider repetition statistics in these surveys as we have estimates for the time spent observing each field, $T_{\mathrm{field}}$. The CHIME data were taken from the first CHIME catalogue and associated papers \citep{CHIME2021, CHIME2023Cat1Reps, Marcote2020_cat1rep, Michilli2023_cat1reps}. The six declination bins listed for CHIME are identical to those of \citet{James2023}, with bounds at $-10.6^{\circ}$, $-4.38^{\circ}$, $6.083^{\circ}$, $38.70^{\circ}$, $78.67^{\circ}$, $85.92^{\circ}$ and $90.0^{\circ}$ respectively. The \ics{} sample is taken from \citet{Shannon2024} and includes FRBs detected until the end of 2023, as this is the point to which we have estimated the pointing times of each field.}
\label{table:rep_surveys}
\begin{tabular}{lccc}
\hline
Survey & No. of FRBs & No. of redshifts & No. of repeaters \\
\hline 
\hline
\CHIMEa{} & 4 & 0 & 0 \\
\CHIMEb{} & 23 & 0 & 0 \\
\CHIMEc{} & 154 & 0 & 3 \\
\CHIMEd{} & 234 & 2 & 12 \\
\CHIMEe{} & 50 & 0 & 1 \\
\CHIMEf{} & 27 & 0 & 0 \\
\hline
\ics{} & 28 & 16 & 0 \\
\hline
\hline
Total & 520 & 18 & 16 \\
\hline
\end{tabular}
\end{center}
\end{table} 

\begin{table}
\begin{center}
\caption{The total number of FRBs and the corresponding number of redshifts that we use in our analysis after the given cuts. These surveys do not consider repetition at all, as there is not sufficient data about the amount of time spent observing each field.}
\label{table:surveys}
\begin{tabular}{lcc}
\hline
Survey & No. of FRBs & No. of redshifts \\
\hline 
\hline
\parkes{} & 16 & 0 \\
\flyseye{} & 25 & 1 \\
DSA & 19 & 14 \\
\FAST{} & 3 & 0 \\
MeerKAT coherent & 10 & 3 \\
MeerKAT incoherent & 4 & 3 \\
\hline
\hline
Total & 86 & 8 \\
\hline
\end{tabular}
\end{center}
\end{table} 

\subsection{CHIME data}
The main focus of this work is to include a repetition model in our full analysis such that we can include the large dataset of CHIME. Here, we only utilise CHIME catalogue 1 data \citep{CHIME2021, CHIME2023Cat1Reps}. Although CHIME catalogue 2 data is available \citep{CHIME2026cat2}, which includes a much larger dataset, the classification of repeating and non-repeating sources has become significantly more blurred. This is due to the high density of sources with relatively low localisation precision, which leads to high potential coincidence associations of single bursts \citep{Cook2026_cat2repeaters}. The catalogue 2 data contains an additional 25 repeaters in a ``gold sample'' and an additional 14 repeaters in a ``silver sample''. The inclusion or exclusion of 14 additional events, and amongst those the choice of which specific events to include as a repeater or non-repeater, would make non-trivial changes to the expected distribution of repeating sources, as it is a significant fraction of the sample. As such, more thought and detailed modelling need to be considered to include this data in a relatively unbiased way.

We use the same CHIME model and data set as \citet{James2023}, which includes 492 FRB sources, of which 16 are repeaters, distributed across six declination bins according to Table \ref{table:rep_surveys}. Of these, repeating FRBs 20180814A, 20180916B and 20190303A have been localised with associated redshifts of 0.068, 0.0337 and 0.064, respectively \citep{Marcote2020_cat1rep, Michilli2023_cat1reps}. As previously mentioned, to avoid observational biases, it is necessary for us to only use redshifts of repeaters in which the number of bursts detected in the given survey ($N_{\mathrm{b,obs}}$) exceeds some minimum burst number ($N_{\mathrm{b,min}}$). This $N_{\mathrm{b,min}}$ limit is chosen such that all repeaters with $N_{\mathrm{b,obs}} > N_{\mathrm{b,min}}$ have associated redshifts. From the first CHIME catalogue, FRBs 20180814A and 20180916B had 11 and 19 bursts, respectively, which are the only two FRBs with more than 3 bursts detected within the observing period. As such, placing a cut such that we only use redshifts from repeaters with 4 or more bursts gives us an unbiased redshift sample, but also means that we do not use the redshift of FRB 20190303A.

\subsection{ASKAP data}
We use all available data from the \parkes{} and \flyseye{} surveys, but use a reduced data set from the \ics{} sample to only include 28 FRBs. Although we do have additional FRBs detected in this survey, we do not have robust logging and estimates of the observation times when they were detected \citep[see][]{Shannon2024}, and hence we cannot use that data for repetition analysis. It is possible to include this data and ignore repetition statistics in this survey as we did previously. However, we choose to focus on analysing repetition in this work. Even without any repeaters identified from this survey, this null detection still provides limits on repetition rates.

CRAFT is a commensal observing program, which means that the frequency configuration and observing fields consistently change. To work around this issue, we approximate all observations with a characteristic frequency of 1.272\,GHz. We also use a single effective pointing time ($T_{\mathrm{eff}}$) of 3.66 days over an effective number of 123 fields. A more detailed description of this modelling is given in \ref{sec:askap_reps}.

\subsection{DSA data} \label{sec:dsa}
We use the sample of 40 localised FRBs discovered by DSA from \citet{Connor2025} and an additional 6 unlocalised FRBs from \citet{Sherman2023}. However, we note that this is not a complete sample of FRBs detected within this observing campaign. To ensure we have an unbiased sample, we must necessarily include all FRBs detected within the survey unless there is a justifiable reason to exclude them in an unbiased way. There were an additional 14 unpublished FRBs detected during these observations, however, there was a technical fault for one preventing any data from being recorded. As such, there are a total of 59 FRBs for which data is available.

As mentioned previously, we can only use redshifts for which $\rm DM_{EG}$ exceeds that of the lowest, unlocalised FRB. When considering all FRBs with $|b| > 20^\circ$, 29 FRBs with 1 associated redshift would be usable in an unbiased way. If we instead consider $|b| > 34^\circ$ for this survey, this reduces the sample to 19 FRBs but allows 14 redshifts to be incorporated. As such, we impose a $|b| > 34^\circ$ restriction only on the DSA sample. 

Of the 19 FRBs, two are in the unpublished sample and are shown in Table \ref{tab:dsa}. FRB `Zeynep' has had standard optical follow-up, but did not have any detectable host galaxy. As such, we exclude the redshifts of FRBs which have $\rm DM_{EG} > DM_{EG, Zeynep} = 610.8$\,\DMunit{} (we note that this value includes \DMhalo{}, but as this is assumed to be identical for all FRBs, it will not affect which redshifts are included).

We are also unable to account for repetition statistics with DSA, as there is no estimate of the total observation time, let alone time per field. Without knowing the amount of time spent observing each field, it is not possible to estimate the expected ratio of single bursts compared to repeaters.

\begin{table}[]
\centering
\begin{tabular}{ccccc}
    Internal name & \DMobs{} & \DMISM{} & SNR & $w$ \\
    & (\DMunit{}) & (\DMunit{}) & & (ms) \\ 
    \hline
    Zeynep & 649.9 & 39.1 & 8.5 & 2.1 \\
    Ramata & 808.8 & 32.8 & 10.1 & 2.1 \\
    \hline
\end{tabular}
\caption{High galactic latitude ($|b| > 34^{\circ}$) FRBs detected by DSA that have not yet been published. Neither have been associated with host galaxies, although both of them have had optical follow-up. \DMG{} is calculated using the \texttt{NE2001} model \citep{Cordes2002} and both FRBs have a width ($w$) of 2 time samples (2.1\,ms).}
\label{tab:dsa}
\end{table}

\subsection{MeerKAT data}
The MeerKAT radio telescope is a precursor of SKA-mid located in South Africa. This radio telescope has detected a significant number of FRBs in blind searches as part of the MeerTRAP (More (meer) TRAnsients and Pulsars) survey, and hence we include this data in our analysis \citep{MeerTRAP2018, Rajwade2022}. It is also more sensitive than Parkes, DSA-110, ASKAP and CHIME, and thus it offers additional constraints by probing a different part of the parameter space.

We model the MeerKAT telescope in two different detection modes described by \citet{Jankowski2023}: A coherent beam (CB) search and an incoherent beam (IB) search. The IB searches use 64 antennas, each with a diameter of 13.5\,m, forming a single IB on the sky with a field of view (FOV) of $\sim 1$\,deg$^2$. This mode has a fluence noise level of 0.34\,Jy\,ms for a 1\,ms burst. The coherent mode instead forms 768 CBs on the sky in a hexagonal pattern, overlapping at the 25\% power point. These beams are formed from the 40 innermost antennas and have a reduced FOV of $\sim 0.19$\,deg$^2$. However, they are $\sim 5$ times more sensitive and hence have a fluence noise level of 0.066\,Jy\,ms. For both modes, the frequency resolution is 0.836\,MHz, the temporal resolution is 0.306\,ms and the SNR threshold is 8. The total DM, DM contribution from the Milky Way interstellar medium (\DMISM{}), central observation frequency ($\nu_c$), bandwidth ($\Delta \nu$), SNR at detection, width ($w$) and redshift of each FRB is presented in Table \ref{tab:MeerKAT}. Although there are multiple observation frequencies, we do not treat them as independent surveys as this would increase the computational requirements, and instead take an average frequency for the coherent and incoherent surveys.

After applying our latitude cut, we use 14 FRBs discovered by MeerKAT with 6 corresponding redshifts from the sample of \citet{PastorMarazuela2025}. These FRBs are listed in Table \ref{tab:MeerKAT}. We include ten FRBs in the coherent survey, and four in the incoherent one. We note that FRBs 20220222C, 20230125D, 20230808F, 20230827E and 20231204B were all detected in both modes, but we only include them in the more sensitive coherent survey to prevent double counting FRBs. We also assume all surveys are independent in our analysis, and hence as the CBs tile the innermost $\sim 0.4$\,deg$^2$ of the IB, the incoherent survey does not include this central region to make the two surveys are independent.

Unlike DSA, there is no $z$ cut above a certain \DMEG{} for MeerTRAP FRBs. The MeerTRAP FRBs without associated redshifts are almost all due to faults, which result in the buffers not being downloaded and hence precise localisations are not possible \citep{PastorMarazuela2025}. The only exceptions to this are FRBs 20230814F, FRB 20231010A and FRB 20231210F. FRB 20230814F has an associated host detected, but has not had spectroscopic follow-up. The photometric redshift obtained was $z=0.29 \pm 0.26$, which we deem to be too uncertain for our analysis. FRB 20231010A had two potential hosts detected with association probabilities of $52.3\%$ and $46.6\%$ respectively. Similarly, FRB 20231210F had 4 potential hosts detected, with the highest having an association probability of $81\%$. As these probabilities do not exceed our arbitrary threshold of $90\%$, we do not include the redshifts. Nevertheless, there are no FRBs which had optical follow-up and did not have any host galaxy detected, and hence we do not need to impose the \DMEG{} cut.


\begin{table*}[]
\centering
\begin{tabular}{ccccccccc}
    TNS & \DMobs{} & \DMISM{} & \hspace{2ex} $\nu_c$ \hspace{2ex} & \hspace{2ex} $\Delta \nu$ \hspace{2ex} & \hspace{2ex} SNR \hspace{2ex} & \hspace{2ex} $w$ \hspace{2ex} & \hspace{2ex} $z$ \hspace{2ex} \\
    & (\DMunit{}) & (\DMunit{}) & (MHz) & (MHz) & & (ms) & \\ 
    \hline
    \hline
    \multicolumn{8}{c}{MeerTRAP Coherent} \\
    \hline
    20220222C & 1071.2 & 56 & 1284 & 770 & 24.2 & 15.9 & 0.853 \\
    20230125D & 640.08 & 88 & 1284 & 770 & 18.28 & 3.67 & 0.3265 \\
    20230306F & 689.5 & 23 & 1284 & 770 & 11.02 & 3.6 & -- \\
    20230413C & 1532.2 & 45 & 816 & 544 & 14.8 & 19.28 & -- \\
    20230613A & 483.51 & 30 & 1284 & 770 & 49.88 & 1.84 & 0.3923 \\
    20230808F & 653.2 & 36 & 1284 & 856 & 35.78 & 7.941 & 0.3472 \\
    20230827E & 1433.7 & 38 & 816 & 544 & 54.8 & 9.6 & -- \\
    20231007C & 2660.4 & 42 & 1284 & 770 & 9.85 & 9.8 & -- \\
    20231010A & 442.59 & 41 & 816 & 544 & 12.77 & 4.82 & -- \\
    20231204B & 1772.1 & 41 & 816 & 544 & 25.0 & 15.0 & -- \\
    20231210F & 720.6 & 32 & 1284 & 770 & 28.3 & 12.86 & -- \\
    \hline
    \multicolumn{8}{c}{MeerTRAP Incoherent} \\
    \hline
    20220224C & 1140.2 & 52 & 1284 & 770 & 11.1 & 46.0 & 0.6271 \\
    20230503E & 483.74 & 88 & 1284 & 770 & 16.87 & 6.7 & -- \\
    20230907D & 1030.79 & 29 & 1284 & 770 & 11.57 & 4.9 & 0.4638 \\
    20231020B & 952.2 & 34 & 1284 & 770 & 16.4 & 16.0 & 0.4775 \\
    \hline
\end{tabular}
\caption{FRBs detected by MeerKAT in both the coherent and incoherent modes \citep[][p. comms with the MeerTRAP team]{PastorMarazuela2025}. FRBs 20220222C, 20230125D, 20230808F, 20230827E and 20231204B were all detected in both modes, but we only include them in the more sensitive coherent survey. This effectively assumes the two surveys are independent, and so the overlapping area is only the coherent survey. The columns correspond to the TNS name, observational DM (\DMobs{}), DM contribution from the Galactic ISM (\DMISM{}) estimated with NE2001 \citep{Cordes2002}, central observational frequency ($\nu_c$), bandwidth ($\Delta \nu$), SNR, width ($w$) and $z$. The FRBs without redshifts were due to instrumental difficulties not allowing the transient buffer data to be downloaded in time, multiple potential hosts or incomplete follow-up to obtain spectroscopic redshifts.}
\label{tab:MeerKAT}
\end{table*}

\section{Results} \label{sec:results}
The results presented in this section necessarily incorporate the observation time per field for repeater surveys (see Section \ref{sec:rep_params}). However, we have previously concluded that for many surveys the total observation time ($T_{\mathrm{obs}}$) is unreliable and gives inconsistent rates between surveys, even when considering different surveys on the same telescope \citep{Shannon2024, Hoffmann2025}. As such, we do not include information on $P(N)$, the probability that $N$ bursts are detected, in the results presented here. Alternative constraints when including $P(N)$ are discussed in \ref{sec:Pn}.

\begin{figure*}
\begin{center}
\includegraphics[width=\linewidth]{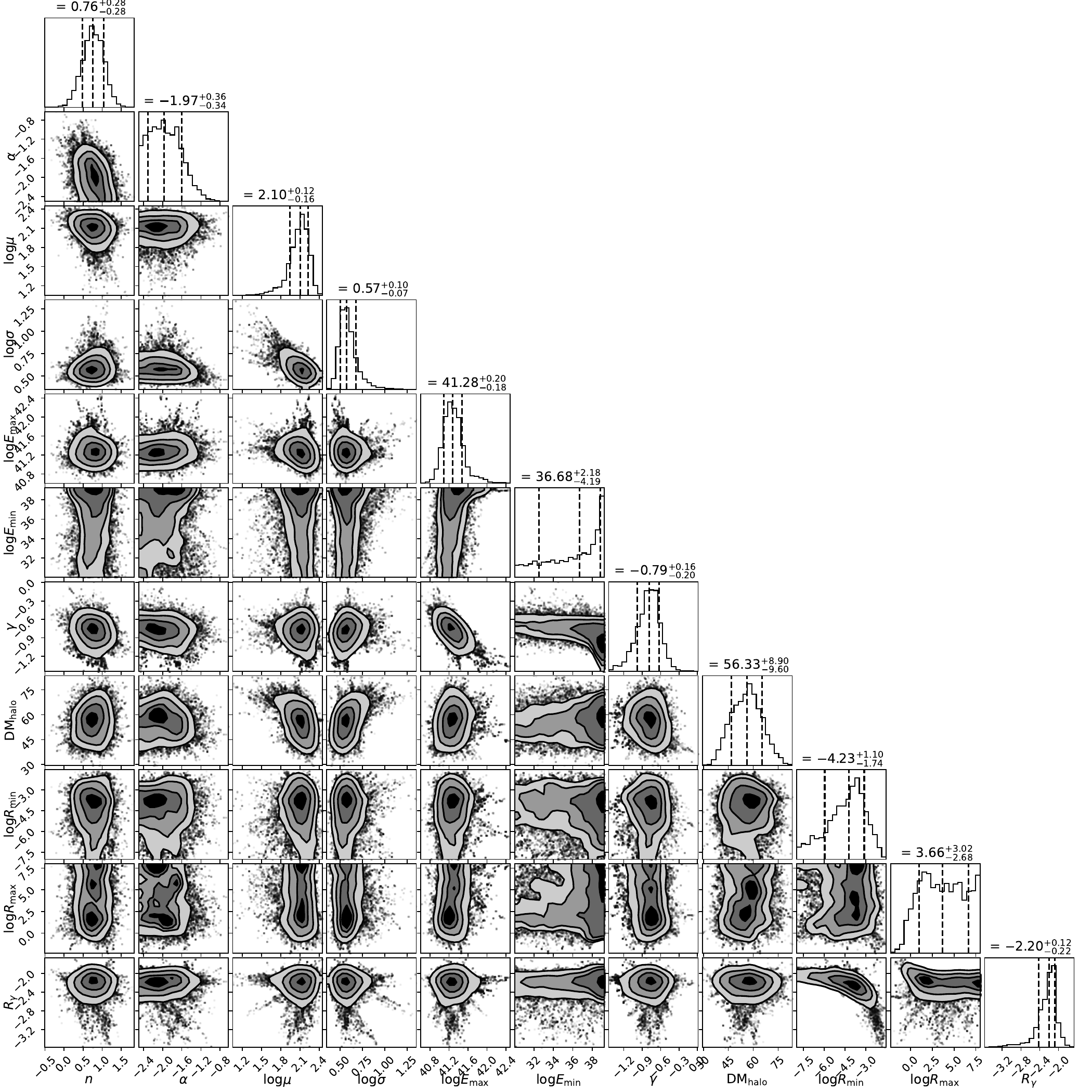}
\caption[]{Results from the MCMC analysis, including repetition parameters. The parameters are described in Table~\ref{table:params}, and the associated priors are also given there. We have excluded $H_0$ from the cornerplot as it is entirely limited by the priors which we assign. This analysis includes repetition statistics from CHIME and ASKAP, which necessarily include the observation time per field, $T_{\mathrm{field}}$. It also includes data from \parkes, \flyseye\, DSA, MeerKAT and FAST; however, no repetition statistics are considered for these surveys due to the lack of robust information on $T_{\mathrm{field}}$. For these surveys without repetition information, we exclude the use of $P(N)$ as we have previously shown that our estimates of the observation time, $T_{\mathrm{obs}}$, give inconsistent results \citep{Hoffmann2025}.}
\label{fig:cornerplot}
\end{center}
\vspace{-3ex}
\end{figure*}

Our primary results are summarised in Figure \ref{fig:cornerplot} and discussed below. The results were generated using 40 walkers running for 2000 steps with a burn-in of 500. Ideally, we would have a significantly larger number of steps, but it is currently not computationally feasible, as even this low number of steps takes $\sim 1$ week to complete. We use a uniform prior on each parameter with limits described in Table \ref{table:params}, following the same priors as \citet{Hoffmann2026}. The priors on $\alpha$ and $H_0$ are informed by existing, independent experimental results.

\begin{table}
\begin{center}
\caption{Limits on the uniform priors used. The repetition parameters \Rmin{}, \Rmax{} and \Rgamma{} describe a power-law distribution of repeater repetition rates. The standard parameters are as follows: \nsfr{} gives the correlation with the cosmic SFR history; $\alpha$ is the slope of the spectral dependence; $\mu_{\mathrm{host}}$ and $\sigma_{\mathrm{host}}$ are the mean and standard deviation of the assumed log-normal distribution of host galaxy DMs; \Emax{} notes the exponential cutoff of the luminosity function (modelled as a Gamma function); \Emin{} is a hard cutoff for the lowest FRB energy; $\gamma$ is the slope of the luminosity function; and $H_0$ is the Hubble constant. \DMhalo{} is in units of \DMunit{}. The host parameters $\mu_{\mathrm{host}}$ and $\sigma_{\mathrm{host}}$ are in units of \DMunit{} in log space, \Emax{} and \Emin{} are in units of ergs and $H_0$ is in units of km$\:$s$^{-1}\:$Mpc$^{-1}$. The limits on $\alpha$ were informed by existing measurements in the literature. The limits on \Emax{} and \Emin{} were chosen as the distributions are uniform on the extrema of these ranges. The limits on $H_0$ represent a 1\,$\sigma$ interval around the \citet{Planck2018} and \citet{SH0ES2021} results.}
\label{table:params}
\begin{tabular}{lcc}
\hline
Parameter & Prior Min & Prior Max \\
\hline 
\hline
\multicolumn{3}{c}{Repetition parameters} \\
\hline
\Rmin{} & -8.0 & -1.0 \\
\Rmax{} & -2.5 & 8.0 \\
\Rgamma{} & -4.0 & -0.5 \\
\hline
\multicolumn{3}{c}{Standard parameters} \\
\hline
\nsfr{} & -2.0 & 6.0 \\
$\alpha$ & -2.5 & -0.5 \\
$\mu_{\mathrm{host}}$ & 1.0 & 3.0 \\
$\sigma_{\mathrm{host}}$ & 0.1 & 1.5 \\
log$_{10}$(\Emax{}) & 38.5 & 42.5 \\
log$_{10}$(\Emin{}) & 30.0 & 40.5 \\
$\gamma$ & -3.0 & 1.0 \\
$H_0$ & 66.9 & 74.08 \\
\DMhalo{} & 10.0 & 200 \\
\hline
\end{tabular}
\end{center}
\end{table} 

\subsection{Predicted number of repeaters and non-repeaters}
Table \ref{table:rates} shows the predicted number of repeaters and non-repeaters for each survey when $P(N)$ information is included only for repetition surveys, and also when it is included for all surveys. Including $P(N)$ only for repetition surveys is a more conservative estimate in which we do not use all of the available data because $T_{\mathrm{obs}}$ has been shown to be unreliable \citep{Shannon2024, Hoffmann2025}. Similarly to previous studies, both cases find the \parkes{} and \flyseye{} surveys overperform while the \ics{} survey underperforms, but is within 1\,$\sigma$ of the expectation assuming Poissonian statistics. The predicted number of repeaters and non-repeaters for \ics{} and CHIME are consistent across all of the declination bins.

When comparing the results including $P(N)$ and excluding it for the non-repetition surveys, there is no significant difference in the results for the repetition surveys, but the predictions for the non-repetition surveys are slightly improved. As such, for this analysis we do not include $P(N)$ for Murriyang (Parkes), \flyseye{} or FAST and only include information of $P(N)$ in the repeater surveys, which effectively determines the ratio of repeaters to non-repeaters. A full analysis including $P(N)$ for the non-repetition surveys is given in \ref{sec:Pn}, but finds no significant differences except in the value of $\alpha$.

\begin{table*}
\begin{center}
\caption{Expected and observed number of repeaters and single bursts in each survey based on the best fit results of Figure \ref{fig:cornerplot}. The expected number of progenitors is shown when only using $T_{\mathrm{field}}$ from the repetition surveys and then when also including $P(N)$ from all surveys. Our main results are those excluding $P(N)$ as the $T_{\mathrm{obs}}$ values are unreliable. The total observation time of DSA and MeerKAT is unknown, and hence cannot be included in this analysis.}
\label{table:rates}
\begin{tabular}{lccccccc}
\hline
Survey & \multicolumn{3}{c}{Repeaters} & \multicolumn{3}{c}{Single bursts} \\
& Exp. No $P(N)$ & Exp. $P(N)$ & Observed & Exp. No $P(N)$ & Exp. $P(N)$ & Observed \\
\hline
\hline
\ics{} & 0.8 & 1.0 & 0 & 33.7 & 42.9 & 28 \\
\CHIMEa{} & 0.1 & 0.1 & 0 & 3.3 & 3.3 & 4 \\
\CHIMEb{} & 0.4 & 0.4 & 0 & 15.5 & 15.6 & 23 \\
\CHIMEc{} & 4.0 & 3.9 & 3 & 155.0 & 156.0 & 151 \\
\CHIMEd{} & 7.3 & 7.2 & 12 & 239.5 & 241.2 & 222 \\
\CHIMEe{} & 1.8 & 1.8 & 1 & 38.0 & 38.2 & 49 \\
\CHIMEf{} & 1.3 & 1.3 & 0 & 19.4 & 19.5 & 27 \\
\hline
\parkes{} & -- & -- & -- & 3.6 & 5.3 & 12 \\
\flyseye{} & -- & -- & -- & 10.0 & 13.3 & 20 \\
\FAST{} & -- & -- & -- & 6.2 & 10.0 & 9 \\
\hline
\end{tabular}
\end{center}
\end{table*}


\subsection{Repetition parameters}

\begin{figure}
    \centering
    \includegraphics[width=\linewidth]{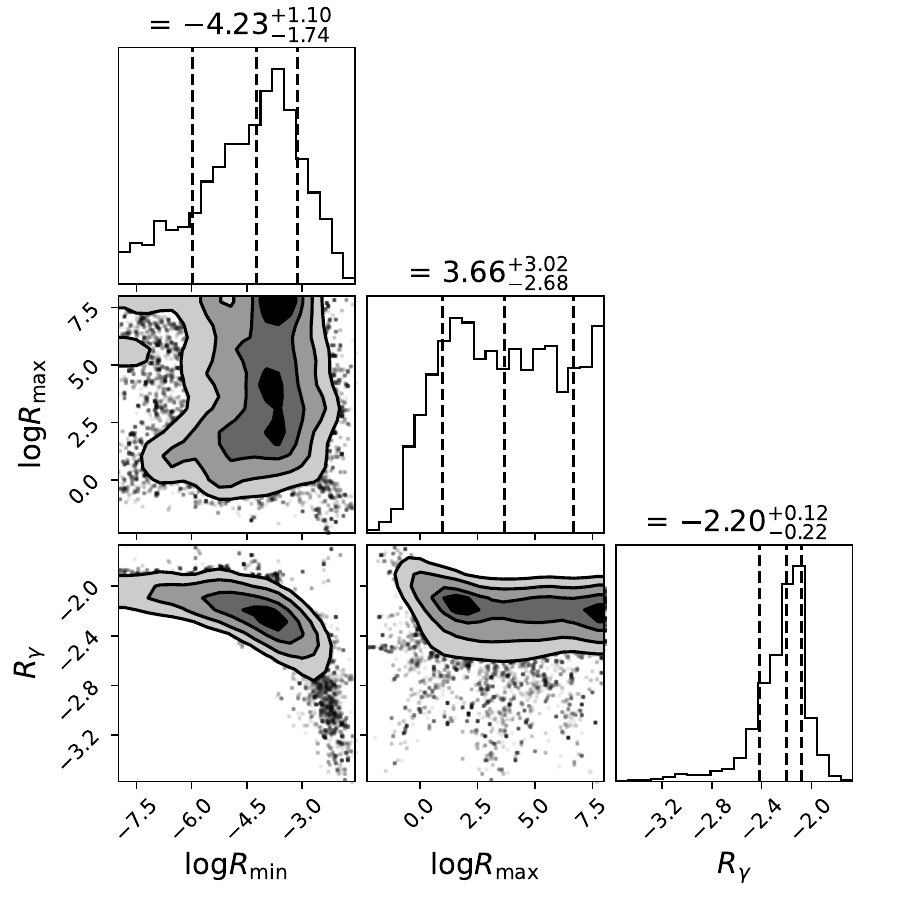}
    \caption{Constraints on the minimum repetition rate \Rmin, the maximum repetition rate \Rmax and the slope of the distribution \Rgamma. These results are a subset of the same data set shown in Figure \ref{fig:cornerplot}.}
    \label{fig:reps}
\end{figure}

An extract of the three repetition parameters \Rmin{}, \Rmax{} and \Rgamma{} is shown in Figure \ref{fig:reps}. These parameters describe a power-law distribution for the repetition rate $R$ above $10^{39}$\,ergs over a 1\,GHz band centred at 1.3\,GHz. Here, \Rmin{} is the minimum repetition rate, \Rmax{} is the maximum repetition rate and \Rgamma{} is the power-law slope of the distribution. For reference, \citet{James2023} estimate the hyperactive repeater FRB 20121102A would have $R = 4$\,d$^{-1}$, accounting for the percentage of time the source is `off' due to its activity window \citep{Rajwade2020, Li2021}.

From this, we constrain \Rgamma{}$=$\Rgammaval{}, \Rmin{}$=$\Rminval{} and \Rmax{}$\gtrsim$\Rmaxval{}. This result is an improvement on the constraints of \citet{James2023} which found \Rgamma{}$=-2.2^{+0.6}_{-0.8}$ and \Rmax > 0.75\,d$^{-1}$, but did not find strong constraints on \Rmin{}. The limits we have on \Rmax{} are consistent with observations from dedicated follow-up campaigns of strong repeaters, which have found repeaters above our lower limit of \Rmaxval{}, although many have unclear activity windows which will affect such an estimate.

We note that \Rgamma{} and \Rmin{} seem to be constrained within the limits of the uniform prior that we choose, and hence we do not expect the choice of prior to have skewed our results. For \Rmin{}, the upper limit is tightly constrained within the prior, with no values exceeding $-1.5$ despite having a limit from the prior of $-1.0$. However, there is an extended tail for the lower limit which extends beyond the limit of $-8.0$ that we artificially impose from the prior. This tail comes from the region of the parameter space where \Rgamma{}$\approx -2$, and the lower limit of \Rmin{} is effectively unconstrained in this region. The prior on \Rmax{} tapers to 0 at the lower limit chosen for the prior, and hence this choice of prior will not have a significant impact on this lower bound. It is clear, however, that we obtain no upper limit on \Rmax{} and hence the values allowed here are limited by the prior chosen.

Interestingly, our results are very close to \Rgamma{}$=-2$. If we assume that the FRB repetition rate is proportional to the angular velocity $\omega$ of a rapidly rotating neutron star, then considering a basic spin-down model and an approximately constant birth rate of FRB progenitors, this gives \Rgamma{}$=-2$, which is consistent with our results.

Both \Rmin{} and \Rmax{} show strong correlations with \Rgamma{}. For steep \Rgamma{} values (\Rgamma{}$\ll -2$), most repeaters would have $R$ close to \Rmin{}, and hence the exact value of \Rmax{} is unconstrained, as these repeaters are so rare. In this instance, if \Rmin{} was too low, all of these repeaters would be observed as single bursts, and thus we would not be able to reproduce the observed number of repeaters. As such, \Rmin{} cannot be too low for steep \Rgamma{} values. If \Rgamma{} is flat (\Rgamma{}$> -2$), \Rmax{} must be low to explain the lack of observed hyperactive repeaters, and hence we see our fits exclude the possibility of a high \Rmax{} alongside a flat \Rgamma{}. This behaviour is similar to what was previously found by \citet{James2023}. Combining the aforementioned constraints, we see that there is a strong preference for \Rgamma{}$=$\Rgammaval{}.

Excluding themselves, the three repetition parameters do not show strong correlations with any other parameters. These three parameters are likely constrained primarily by the ratio of repeaters to non-repeaters, which will not be heavily affected by the other parameters of our model. As such, this suggests we can fit repetition parameters independently of the other population parameters.

From our constraints on the repetition parameters, we then determine the rate of FRBs ($\Phi$) exceeding $10^{39}$\,erg in a GHz bandwidth centred at 1.3\,GHz in the emission frame at $z=0$. From this, we determine the total number of repeaters ($C_r$) at $z=0$ to be log$_{10}C_r = -3.9^{+1.5}_{-0.9}$\,Mpc$^{-3}$. The uncertainty was calculated by taking a 68\% confidence interval around the median value of the log$_{10}C_r$ distribution, constructed from 1400 realisations of the repetition parameters in the MCMC chains. As we assume all apparent non-repeaters are intrinsically repeaters, this corresponds to the total density of FRB progenitors. The highest number of predicted progenitors occurs around \Rgamma{}$\approx-2$, as values of \Rgamma{} flatter than this prefer a small number of strong repeaters and steeper values of \Rgamma{} require high \Rmin{} values (and thereby more repeaters) to prevent an excess of apparent non-repeaters. As such, our preferred value of \Rgamma{} residing near the $-2$ mark causes large uncertainties in the value of $C_r$. Magnetar fields are expected to decay on timescales of order $10^{4}$\,yr \citep{Colpi2000}, and under this assumption we predict birth rates of $1 \sim 400$\,Gpc$^{-3}$\,yr$^{-1}$. This is now inconsistent with the birth rates of core-collapse supernovae at the $\sim 2\,\sigma$ level, which have birth rates of $10^5$\,Gpc$^{-3}$\,yr$^{-1}$ \citep{Taylor2014}. This does not strongly exclude such a model, but suggests that not all young magnetars would be detected as FRBs.

\subsection{Other parameter constraints}
Excluding the three new repetition parameters, we have also extended our data set to include a large number of unlocalised FRBs from CHIME, more localised FRBs from DSA, and the MeerKAT sample of FRBs. CHIME has the largest sample of unlocalised FRBs available and hence can lend more statistical constraining power despite not having localisations. Additionally, CHIME observes at a lower frequency than all other surveys, which gives more constraining power on the slope of the spectral distribution ($\alpha$). This does come with the caveat that we must necessarily use $P(N)$ alongside surveys of different frequency bands to constrain $\alpha$. While we must incorporate $P(N)$ for repetition surveys, the reliability of such an estimate is unclear. DSA has a large number of localised FRBs now, with a comparable number of usable redshifts to the \ics{} survey. As such, this greatly increases our statistical constraining power for all considered parameters. Similar to FAST, MeerKAT is a much more sensitive instrument than those previously used in our analysis \citep[see][]{Caleb2025z2} and also includes localised FRBs, allowing us to probe further into the Universe and thereby help constrain parameters such as \nsfr{}.

Compared to our previous results \citep{Hoffmann2026}, the precision of our constraints improves by a factor of $\sim 2$ for most of the parameters while remaining around the same central values.

We place a prior on $\alpha$ of $-2.5 < \alpha < -0.5$ based on observational results of FRBs found in previous studies \citep{Macquart2019, shin2023, Cui2025}. This parameter is primarily constrained by comparing the number of FRBs detected by telescopes observing in different frequency bands. Previously, we could not further constrain this parameter in the broad window allowed, as we only used surveys covering similar frequency ranges and which had unreliable $T_{\mathrm{obs}}$ measurements. With the inclusion of a large number of FRBs from CHIME, which operates at lower frequencies of 400\,MHz -- 800\,MHz, this increases the frequency lever arm we have to constrain the spectral distribution of FRBs. Additionally, we must incorporate some $P(N)$ information now from the repetition surveys, as it is necessary when doing repetition analysis. As such, we naturally expect a stronger constraint on $\alpha$ than before. We now no longer simply return the prior and obtain stronger constraints of $\alpha = -1.97^{+0.36}_{-0.34}$. We obtain a particularly strong upper limit, but are still bounded by the prior for steep values of $\alpha$. 

In \ref{sec:Pn}, we include the estimates of $T_{\mathrm{obs}}$ that we have for \parkes{}, \flyseye{} and \FAST{} and instead find $\alpha = -1.55^{+0.35}_{-0.29}$, with good constraints on the steep-end slope. At first glance, these results are exactly consistent with the original findings of \citet{Macquart2019}, which found $\alpha = -1.5^{+0.2}_{-0.3}$. However, here we use a rate interpretation of $\alpha$, while \citet{Macquart2019} used a spectral index interpretation and so we would actually expect a value of $\alpha = -0.65^{+0.2}_{-0.3}$ \citep[as argued by][]{james2022b} for consistent results. 

The FRB source evolution parameter showed the largest changes from our previous analysis \citep{Hoffmann2026}. Previously it was found to be \nsfr{}$= 2.88^{+0.83}_{-0.76}$, but now is seen to be $0.76^{+0.28}_{-0.28}$. These values are more similar to the constraints that were previously found from \citep{James2022a} and \citet{Hoffmann2025} and are once again consistent with FRBs tracing the star-formation history of the Universe or being slightly damped. We note that \nsfr{} is strongly correlated with $\alpha$ and hence the improved constraints on $\alpha$ are likely the reason for the large changes in \nsfr{}. Alternatively, we also include results from MeerKAT now, which is more sensitive than the other telescopes that have localised FRBs. As such, it gives much more leverage in understanding the evolution of FRBs further back in the history of the Universe. Additionally, as we gain more discerning power on \nsfr{}, it is also possible to consider alternative models for the source evolution. There are many suggestions in the literature promoting time-delayed models rather than the naive approximation of scaling directly with the star formation history of the Universe, which we plan to implement in a future iteration of \texttt{zDM} \citep[e.g.][]{Zhang2022_timedelay, Hashimoto2022_nsfr, Qiang2022_nsfr, Zhang2023_nsfr, Chen2024_nsfr, Om2025_nsfr, Zhang2026_timedelay}.

Similar to previous analyses, we still find the lower limit of \Emin{} is unconstrained \citep{Hoffmann2025} and the code prefers as high a value as possible in order to maximise the likelihoods of each data point. To really constrain \Emin{}, we would need to find a deficit of low-energy FRBs which are most likely to come from the nearby Universe. Sensitive telescopes with a relatively small FOV such as MeerKAT are unlikely to fill such a gap, as these FRBs are only detectable in the near Universe where the visible volume from such an instrument is very low, thereby making it difficult to detect a statistically significant deficit. Conversely, instruments which are not sensitive enough to probe this faint region of the luminosity function despite having a large FOV are also not sufficient. We hope that as CHIME obtains more localisations with their new outriggers, or as more sensitive survey instruments such as the (CRA)FT (CO)herent upgrade (CRACO) and new iterations of DSA build larger samples, they will be able to lend more insight into this.

Lastly, we note that the uncertainty on \DMhalo{} reduced by a factor of $\sim 3$ in comparison to \citet{Hoffmann2026} which found \DMhalo{}$= 68^{+27}_{-24}$\,\DMunit{}. These results are much more conclusive and are in surprisingly good agreement with the estimate of 50\,\DMunit{} that was used previously \citep{Prochaska2019a}.

\subsection{CHIME $z$-DM distribution}
\begin{figure}
    \centering
    \includegraphics[width=\linewidth]{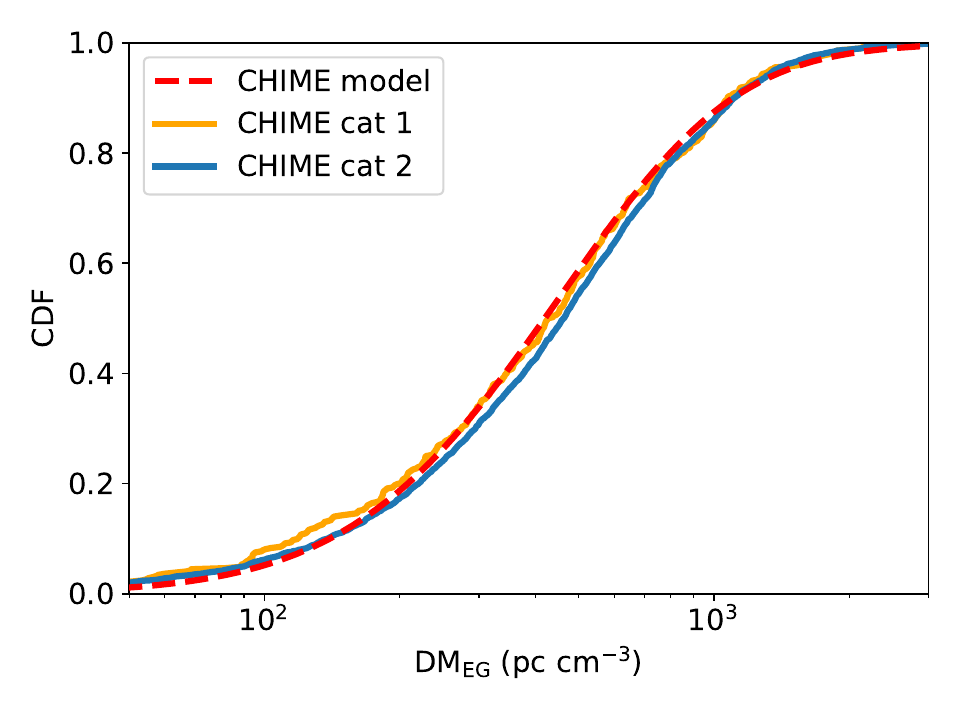}
    \caption{Cumulative \DMEG{} distribution of CHIME FRBs. The red dashed line shows the predicted distribution from our best-fit parameters. The orange line shows the cumulative CHIME catalogue 1 data and the blue line shows the catalogue 2 data. The KS-test returns a $p$-value of 0.9 for the catalogue 1 data and $1.8 \times 10^{-4}$ for the catalogue 2 data. We have not included uncertainties here, but there are additional uncertainties when calculating \DMEG{} from \DMISM{} and \DMhalo{} estimates.}
    \label{fig:CHIMEdm}
\end{figure}

Figure \ref{fig:CHIMEdm} shows the cumulative \DMEG{} distribution of CHIME FRBs in comparison to our predicted distribution from the best-fit parameters of this analysis. The Kolmogorov-Smirnov (KS) test returns a test statistic of 0.037 corresponding to a $p$-value of 0.9, indicating surprisingly good agreement for such a large number of data points. We do not consider uncertainty in \DMEG{}, which would be present due to inaccuracies in \DMISM{} \citep[which can be as high as 50\%;][]{Schnitzeler2012} and \DMhalo{} (which is of order 10\,\DMunit{} from Figure \ref{fig:cornerplot}). Our fit is an improvement over \citet{James2023}, where the author found that they could approximately fit the CHIME catalogue 1 \DMEG{} distribution for a range of different repetition parameters. Prior to this, \citet{Chawla2022} simultaneously fit the DM and scattering distributions of the first CHIME catalogue but did not find a model which statistically favoured the data. Later, \citet{shin2023} created a population model for the CHIME catalogue 1 sample which was consistent with the data and is comparable to our results. These results specifically fit the CHIME catalogue 1 distribution, while ours also includes other surveys. As such, it is reassuring to retain consistency even between surveys.

We also show a comparison to the CHIME catalogue 2 dataset \citep{CHIME2026cat2}, which was not included in the fitting process. Our predicted model visually matches the DM distribution from the extended sample well, but returns a KS test statistic of 0.05 corresponding to a $p$-value of $1.8 \times 10^{-4}$. When comparing the catalogue 1 and 2 data directly, the KS test returns a test statistic of 0.04 with a $p$-value of 0.7.

\begin{figure}
    \centering
    \includegraphics[width=\linewidth]{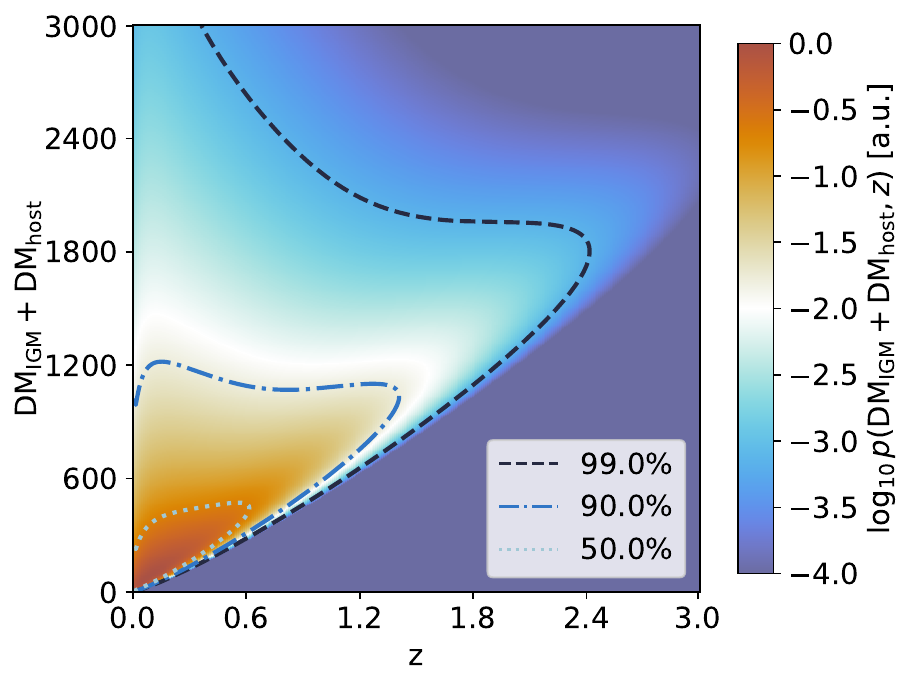}
    \caption{Predicted $z$--DM distribution of CHIME FRBs using the best-fit parameters from this analysis. The colour bar represents the probability of an FRB progenitor being located at a given $z$ and DM value. The contours encase regions containing 50\%, 90\% and 99\% of the total probability density.}
    \label{fig:CHIMEzDM}
\end{figure}

With the same best-fit values, we predict that 7\% of all CHIME-detected FRBs will have $z>1$. As CHIME have demonstrated that they are now able to localise bursts much more precisely with the new CHIME outriggers \citep{CHIME2025Outriggers}, such localisations will be feasible going forward. Figure \ref{fig:CHIMEzDM} shows the full $z$-DM distribution we predict. In this interpretation, the colour bar represents the density of FRB progenitors, accounting for both repeaters and non-repeaters.

\subsection{Model limitations} \label{sec:limitations}
While our results show remarkable self-consistency in reproducing the DM distribution of CHIME and the expected number of repeaters and non-repeaters across \ics{} and each CHIME declination bin, there are a number of limitations at present. Our model assumes that all FRBs are from the same intrinsic population and hence can be described by one set of population parameters. That is, both repeaters and non-repeaters are drawn from the same global energy distribution, have the same source evolution and the same global spectral behaviour. However, distinctions between the widths, spectral behaviour and burst morphology between repeaters and non-repeaters have been observed previously \citep[e.g.][]{CHIME2021, Pleunis2021morphology}.

Additionally, when calculating $P(N_{\mathrm{bursts}})$ for each repeater, we assume Poissonian statistics, given this FRB has a fixed repetition rate, $R$. Observationally, repeaters have been shown to have activity windows \citep[e.g.][]{Chime2020repetition, Rajwade2020, Cruces2021, Braga2025}, in which they are inactive for extended periods of time and then have periods of active phases. As such, simply assuming the number of bursts detected from a repeater is dependent on a single number has obvious limitations, as it will be highly dependent on the phase of the repeater during the observational duration. Additionally, they exhibit non-Poissonian clustering on sub-arcsecond to minute timescales which we do not consider. We aim to revise the treatment of this within \texttt{zDM} once more is known of the long-term behaviour of repeating FRBs.

\section{Conclusion} \label{sec:conclusion}
In this work, we present the first fit of FRB population parameters that includes a repetition model. This model assumes all FRBs are intrinsic repeaters, and models the repetition rates as a simple power-law distribution. We incorporate the data from Murriyang (Parkes), ASKAP, FAST, DSA, MeerKAT and CHIME to significantly increase our overall sample size and overall number of FRBs with associated $z$ values. We model repetition in \ics{} and CHIME, while treating the other surveys as we did previously. We only use CHIME catalogue 1 data as we cannot robustly determine which FRBs are repeaters in catalogue 2. This data will be included in future analyses, but more thought needs to be given on how to account for uncertainty in repeater identification.

We constrain the minimum repetition rate of bursts above $10^{39}$\,erg to \Rmin{}$=$\Rminval{}, the maximum repetition rate to \Rmax{}$\gtrsim$\Rmaxval{} and the power-law index to \Rgamma{}$=$\Rgammaval{}. These results are consistent with previous work from \citet{James2023} but more precise. We find that the repetition parameters show no significant correlation with any of the other FRB population parameters. Our best-fit constraints show good self-consistency, being able to reproduce the CHIME DM distribution for both catalogue 1 and catalogue 2 despite catalogue 2 not being part of the fit. We also produce reasonable predictions of the number of repeaters and non-repeaters detected in \ics{} and each CHIME declination bin. We predict an FRB birth rate of $1 \sim 400$\,Gpc$^{-3}$\,yr$^{-1}$, which is $2\,\sigma$ lower than the predicted birth rates of magnetars from core-collapse supernovae.

The inclusion of these larger datasets has also shown significant improvements in the precision of our results, not just for the repetition parameters but also including all other FRB population parameters. As CHIME, DSA, MeerKAT and CRACO are predicted to detect an increasing number of localised FRBs in the coming years, we hope that population analyses similar to this work are able to tell us more about the intrinsic nature of FRBs, and also allow us to more accurately model the FRB population so that they can also be used in precision cosmology.

\begin{acknowledgement}
This work was performed on the OzSTAR national facility at Swinburne University of Technology. The OzSTAR programme receives funding in part from the Astronomy National Collaborative Research Infrastructure Strategy (NCRIS) allocation provided by the Australian Government.

This scientific work uses data obtained from Inyarrimanha Ilgari Bundara, the CSIRO Murchison Radio-astronomy Observatory. We acknowledge the Wajarri Yamaji as the Traditional Owners and native title holders of the Observatory site. CSIRO’s ASKAP radio telescope is part of the Australia Telescope National Facility. The operation of ASKAP is funded by the Australian Government with support from the National Collaborative Research Infrastructure Strategy. ASKAP uses the resources of the Pawsey Supercomputing Research Centre. The establishment of ASKAP, Inyarrimanha Ilgari Bundara, the CSIRO Murchison Radio-astronomy Observatory and the Pawsey Supercomputing Research Centre are initiatives of the Australian Government, with support from the Government of Western Australia and the Science and Industry Endowment Fund.

\end{acknowledgement}

\paragraph{Funding Statement}
This research was supported by an Australian Government Research Training Program (RTP) Scholarship.
CWJ acknowledges support through Australian Research Council (ARC) Discovery Project (DP) DP210102103.
IPM acknowledges funding for this work from the European Research Council (ERC) under the European Union’s Horizon 2020 research and innovation programme (`EuroFlash’; Grant agreement No. 101098079).

\paragraph{Competing Interests}
None.

\paragraph{Data Availability Statement}
The code and data used to produce our results can be found at \href{https://github.com/FRBs/zdm}{https://github.com/FRBs/zdm}.

\printendnotes

\printbibliography 

@ARTICLE{Zhang2026_timedelay,
       author = {{Zhang}, Zi-Liang and {Zhang}, Bing},
        title = "{Evidence for a Delayed Progenitor Population for CHIME non-repeating Fast Radio Bursts using a Self-Consistent Forward and Backward Inference Framework}",
      journal = {arXiv e-prints},
         year = 2026,
        month = jul,
          eid = {arXiv:2607.04792},
        pages = {arXiv:2607.04792},
          doi = {10.48550/arXiv.2607.04792},
archivePrefix = {arXiv},
       eprint = {2607.04792},
 primaryClass = {astro-ph.HE},
       adsurl = {https://ui.adsabs.harvard.edu/abs/2026arXiv260704792Z}
}

@ARTICLE{Om2025_nsfr,
       author = {{Gupta}, Om and {Beniamini}, Paz and {Kumar}, Pawan and {Finkelstein}, Steven L.},
        title = "{The Cosmic Evolution of Fast Radio Bursts Inferred from the CHIME/FRB Baseband Catalog 1}",
      journal = {\apj},
         year = 2025,
        month = jun,
       volume = {986},
       number = {1},
          eid = {100},
        pages = {100},
          doi = {10.3847/1538-4357/add14c},
archivePrefix = {arXiv},
       eprint = {2501.09810},
 primaryClass = {astro-ph.HE},
       adsurl = {https://ui.adsabs.harvard.edu/abs/2025ApJ...986..100G}
}

@ARTICLE{Hashimoto2022_nsfr,
       author = {{Hashimoto}, Tetsuya and {Goto}, Tomotsugu and {Chen}, Bo Han and {Ho}, Simon C.-C. and {Hsiao}, Tiger Y.-Y. and {Wong}, Yi Hang Valerie and {On}, Alvina Y.~L. and {Kim}, Seong Jin and {Kilerci-Eser}, Ece and {Huang}, Kai-Chun and {Santos}, Daryl Joe D. and {Yamasaki}, Shotaro},
        title = "{Energy functions of fast radio bursts derived from the first CHIME/FRB catalogue}",
      journal = {\mnras},
         year = 2022,
        month = apr,
       volume = {511},
       number = {2},
        pages = {1961-1976},
          doi = {10.1093/mnras/stac065},
archivePrefix = {arXiv},
       eprint = {2201.03574},
 primaryClass = {astro-ph.HE},
       adsurl = {https://ui.adsabs.harvard.edu/abs/2022MNRAS.511.1961H}
}

@ARTICLE{Qiang2022_nsfr,
       author = {{Qiang}, Da-Chun and {Li}, Shu-Ling and {Wei}, Hao},
        title = "{Fast radio burst distributions consistent with the first CHIME/FRB catalog}",
      journal = {\jcap},
         year = 2022,
        month = jan,
       volume = {2022},
       number = {1},
          eid = {040},
        pages = {040},
          doi = {10.1088/1475-7516/2022/01/040},
archivePrefix = {arXiv},
       eprint = {2111.07476},
 primaryClass = {astro-ph.HE},
       adsurl = {https://ui.adsabs.harvard.edu/abs/2022JCAP...01..040Q}
}

@ARTICLE{Zhang2023_nsfr,
       author = {{Zhang}, Zi-Liang and {Yu}, Yun-Wei and {Cao}, Xiao-Feng},
        title = "{Diverse origins for non-repeating fast radio bursts: Rotational radio transient sources and cosmological compact binary merger remnants}",
      journal = {\aap},
         year = 2023,
        month = jul,
       volume = {675},
          eid = {A66},
        pages = {A66},
          doi = {10.1051/0004-6361/202245511},
archivePrefix = {arXiv},
       eprint = {2303.14695},
 primaryClass = {astro-ph.HE},
       adsurl = {https://ui.adsabs.harvard.edu/abs/2023A&A...675A..66Z}
}

@ARTICLE{Chen2024_nsfr,
       author = {{Chen}, J.~H. and {Jia}, X.~D. and {Dong}, X.~F. and {Wang}, F.~Y.},
        title = "{The Formation Rate and Luminosity Function of Fast Radio Bursts}",
      journal = {\apjl},
         year = 2024,
        month = oct,
       volume = {973},
       number = {2},
          eid = {L54},
        pages = {L54},
          doi = {10.3847/2041-8213/ad7b39},
archivePrefix = {arXiv},
       eprint = {2406.03672},
 primaryClass = {astro-ph.HE},
       adsurl = {https://ui.adsabs.harvard.edu/abs/2024ApJ...973L..54C}
}

@ARTICLE{Zhang2022_timedelay,
       author = {{Zhang}, Rachel C. and {Zhang}, Bing},
        title = "{The CHIME Fast Radio Burst Population Does Not Track the Star Formation History of the Universe}",
      journal = {\apjl},
         year = 2022,
        month = jan,
       volume = {924},
       number = {1},
          eid = {L14},
        pages = {L14},
          doi = {10.3847/2041-8213/ac46ad},
archivePrefix = {arXiv},
       eprint = {2109.07558},
 primaryClass = {astro-ph.HE},
       adsurl = {https://ui.adsabs.harvard.edu/abs/2022ApJ...924L..14Z}
}

@ARTICLE{James2026_path_zdm,
       author = {{James}, C.~W. and {Andersen}, B.~C. and {Marnoch}, L. and {Hoffmann}, J.~L. and {Loudas}, N. and {Prochaska}, J.~X. and {Ryder}, S.~D. and {Woodland}, M.},
        title = "{Through a glass, darkly: a combined framework for estimating fast radio burst host galaxy and population properties in an era of uncertain host identification}",
      journal = {arXiv e-prints},
         year = 2026,
        month = sep,
          eid = {arXiv:2609.04765},
        pages = {arXiv:2609.04765},
archivePrefix = {arXiv},
       eprint = {2609.04765},
 primaryClass = {astro-ph.HE},
       adsurl = {https://ui.adsabs.harvard.edu/abs/2026arXiv260904765J}
}

@ARTICLE{Marcote2020_cat1rep,
       author = {{Marcote}, B. and {Nimmo}, K. and {Hessels}, J.~W.~T. and {Tendulkar}, S.~P. and {Bassa}, C.~G. and {Paragi}, Z. and {Keimpema}, A. and {Bhardwaj}, M. and {Karuppusamy}, R. and {Kaspi}, V.~M. and {Law}, C.~J. and {Michilli}, D. and {Aggarwal}, K. and {Andersen}, B. and {Archibald}, A.~M. and {Bandura}, K. and {Bower}, G.~C. and {Boyle}, P.~J. and {Brar}, C. and {Burke-Spolaor}, S. and {Butler}, B.~J. and {Cassanelli}, T. and {Chawla}, P. and {Demorest}, P. and {Dobbs}, M. and {Fonseca}, E. and {Giri}, U. and {Good}, D.~C. and {Gourdji}, K. and {Josephy}, A. and {Kirichenko}, A. Yu. and {Kirsten}, F. and {Landecker}, T.~L. and {Lang}, D. and {Lazio}, T.~J.~W. and {Li}, D.~Z. and {Lin}, H.-H. and {Linford}, J.~D. and {Masui}, K. and {Mena-Parra}, J. and {Naidu}, A. and {Ng}, C. and {Patel}, C. and {Pen}, U.-L. and {Pleunis}, Z. and {Rafiei-Ravandi}, M. and {Rahman}, M. and {Renard}, A. and {Scholz}, P. and {Siegel}, S.~R. and {Smith}, K.~M. and {Stairs}, I.~H. and {Vanderlinde}, K. and {Zwaniga}, A.~V.},
        title = "{A repeating fast radio burst source localized to a nearby spiral galaxy}",
      journal = {\nat},
         year = 2020,
        month = jan,
       volume = {577},
       number = {7789},
        pages = {190-194},
          doi = {10.1038/s41586-019-1866-z},
archivePrefix = {arXiv},
       eprint = {2001.02222},
 primaryClass = {astro-ph.HE},
       adsurl = {https://ui.adsabs.harvard.edu/abs/2020Natur.577..190M}
}

@ARTICLE{Michilli2023_cat1reps,
       author = {{Michilli}, Daniele and {Bhardwaj}, Mohit and {Brar}, Charanjot and {Gaensler}, B.~M. and {Kaspi}, Victoria M. and {Kirichenko}, Aida and {Masui}, Kiyoshi W. and {Mckinven}, Ryan and {Ng}, Cherry and {Patel}, Chitrang and {Sand}, Ketan R. and {Scholz}, Paul and {Shin}, Kaitlyn and {Siegel}, Seth R. and {Stairs}, Ingrid and {Cassanelli}, Tomas and {Cook}, Amanda M. and {Dobbs}, Matt and {Dong}, Fengqiu Adam and {Fonseca}, Emmanuel and {Ibik}, Adaeze and {Kaczmarek}, Jane and {Leung}, Calvin and {Pearlman}, Aaron B. and {Petroff}, Emily and {Pleunis}, Ziggy and {Rafiei-Ravandi}, Masoud and {Sanghavi}, Pranav and {Shaw}, J. Richard and {Tendulkar}, Shriharsh P.},
        title = "{Subarcminute Localization of 13 Repeating Fast Radio Bursts Detected by CHIME/FRB}",
      journal = {\apj},
         year = 2023,
        month = jun,
       volume = {950},
       number = {2},
          eid = {134},
        pages = {134},
          doi = {10.3847/1538-4357/accf89},
archivePrefix = {arXiv},
       eprint = {2212.11941},
 primaryClass = {astro-ph.HE},
       adsurl = {https://ui.adsabs.harvard.edu/abs/2023ApJ...950..134M}
}

@ARTICLE{FAST,
       author = {{Nan}, Rendong},
        title = "{Five hundred meter aperture spherical radio telescope (FAST)}",
      journal = {Science in China: Physics, Mechanics and Astronomy},
         year = 2006,
        month = mar,
       volume = {49},
       number = {2},
        pages = {129-148},
          doi = {10.1007/s11433-006-0129-9},
       adsurl = {https://ui.adsabs.harvard.edu/abs/2006ScChG..49..129N}
}

@INPROCEEDINGS{CHIME,
       author = {{Bandura}, Kevin and {Addison}, Graeme E. and {Amiri}, Mandana and {Bond}, J. Richard and {Campbell-Wilson}, Duncan and {Connor}, Liam and {Cliche}, Jean-Fran{\c{c}}ois and {Davis}, Greg and {Deng}, Meiling and {Denman}, Nolan and {Dobbs}, Matt and {Fandino}, Mateus and {Gibbs}, Kenneth and {Gilbert}, Adam and {Halpern}, Mark and {Hanna}, David and {Hincks}, Adam D. and {Hinshaw}, Gary and {H{\"o}fer}, Carolin and {Klages}, Peter and {Landecker}, Tom L. and {Masui}, Kiyoshi and {Mena Parra}, Juan and {Newburgh}, Laura B. and {Pen}, Ue-li and {Peterson}, Jeffrey B. and {Recnik}, Andre and {Shaw}, J. Richard and {Sigurdson}, Kris and {Sitwell}, Mike and {Smecher}, Graeme and {Smegal}, Rick and {Vanderlinde}, Keith and {Wiebe}, Don},
        title = "{Canadian Hydrogen Intensity Mapping Experiment (CHIME) pathfinder}",
    booktitle = {Ground-based and Airborne Telescopes V},
         year = 2014,
       editor = {{Stepp}, Larry M. and {Gilmozzi}, Roberto and {Hall}, Helen J.},
       series = {Society of Photo-Optical Instrumentation Engineers (SPIE) Conference Series},
       volume = {9145},
        month = jul,
          eid = {914522},
        pages = {914522},
          doi = {10.1117/12.2054950},
archivePrefix = {arXiv},
       eprint = {1406.2288},
 primaryClass = {astro-ph.IM},
       adsurl = {https://ui.adsabs.harvard.edu/abs/2014SPIE.9145E..22B}
}

@ARTICLE{Baptista2023,
       author = {{Baptista}, Jay and {Prochaska}, J. Xavier and {Mannings}, Alexandra G. and {James}, C.~W. and {Shannon}, R.~M. and {Ryder}, Stuart D. and {Deller}, A.~T. and {Scott}, Danica R. and {Glowacki}, Marcin and {Tejos}, Nicolas},
        title = "{Measuring the Variance of the Macquart Relation in z-DM Modeling}",
      journal = {arXiv e-prints},
         year = 2023,
        month = may,
          eid = {arXiv:2305.07022},
        pages = {arXiv:2305.07022},
          doi = {10.48550/arXiv.2305.07022},
archivePrefix = {arXiv},
       eprint = {2305.07022},
 primaryClass = {astro-ph.CO},
       adsurl = {https://ui.adsabs.harvard.edu/abs/2023arXiv230507022B}
}

@ARTICLE{James2023,
       author = {{James}, C.~W.},
        title = "{Modelling repetition in zDM: A single population of repeating fast radio bursts can explain CHIME data}",
      journal = {\pasa},
         year = 2023,
        month = dec,
       volume = {40},
          eid = {e057},
        pages = {e057},
          doi = {10.1017/pasa.2023.51},
archivePrefix = {arXiv},
       eprint = {2306.17403},
 primaryClass = {astro-ph.HE},
       adsurl = {https://ui.adsabs.harvard.edu/abs/2023PASA...40...57J}
}

@ARTICLE{Macquart2020,
       author = {{Macquart}, J.~-P. and {Prochaska}, J.~X. and {McQuinn}, M. and {Bannister}, K.~W. and {Bhandari}, S. and {Day}, C.~K. and {Deller}, A.~T. and {Ekers}, R.~D. and {James}, C.~W. and {Marnoch}, L. and {Os{\l}owski}, S. and {Phillips}, C. and {Ryder}, S.~D. and {Scott}, D.~R. and {Shannon}, R.~M. and {Tejos}, N.},
        title = "{A census of baryons in the Universe from localized fast radio bursts}",
      journal = {\nat},
         year = 2020,
        month = may,
       volume = {581},
       number = {7809},
        pages = {391-395},
          doi = {10.1038/s41586-020-2300-2},
archivePrefix = {arXiv},
       eprint = {2005.13161},
 primaryClass = {astro-ph.CO},
       adsurl = {https://ui.adsabs.harvard.edu/abs/2020Natur.581..391M}
}

@ARTICLE{james2022,
       author = {{James}, C.~W. and {Prochaska}, J.~X. and {Macquart}, J.~-P. and {North-Hickey}, F.~O. and {Bannister}, K.~W. and {Dunning}, A.},
        title = "{The z-DM distribution of fast radio bursts}",
      journal = {\mnras},
         year = 2022,
        month = feb,
       volume = {509},
       number = {4},
        pages = {4775-4802},
          doi = {10.1093/mnras/stab3051},
archivePrefix = {arXiv},
       eprint = {2101.08005},
 primaryClass = {astro-ph.HE},
       adsurl = {https://ui.adsabs.harvard.edu/abs/2022MNRAS.509.4775J}
}

@ARTICLE{Chime2020repetition,
       author = {{CHIME/FRB Collaboration} and {Amiri}, M. and {Andersen}, B.~C. and {Band
        ura}, K.~M. and {Bhardwaj}, M. and {Boyle}, P.~J. and {Brar}, C. and
         {Chawla}, P. and {Chen}, T. and {Cliche}, J.~F. and {Cubranic}, D. and
         {Deng}, M. and {Denman}, N.~T. and {Dobbs}, M. and {Dong}, F.~Q. and {Fand
        ino}, M. and {Fonseca}, E. and {Gaensler}, B.~M. and {Giri}, U. and
         {Good}, D.~C. and {Halpern}, M. and {Hessels}, J.~W.~T. and
         {Hill}, A.~S. and {H{\"o}fer}, C. and {Josephy}, A. and {Kania}, J.~W. and
         {Karuppusamy}, R. and {Kaspi}, V.~M. and {Keimpema}, A. and
         {Kirsten}, F. and {Landecker}, T.~L. and {Lang}, D.~A. and {Leung}, C. and
         {Li}, D.~Z. and {Lin}, H. -H. and {Marcote}, B. and {Masui}, K.~W. and
         {McKinven}, R. and {Mena-Parra}, J. and {Merryfield}, M. and
         {Michilli}, D. and {Milutinovic}, N. and {Mirhosseini}, A. and
         {Naidu}, A. and {Newburgh}, L.~B. and {Ng}, C. and {Nimmo}, K. and
         {Paragi}, Z. and {Patel}, C. and {Pen}, U. -L. and
         {Pinsonneault-Marotte}, T. and {Pleunis}, Z. and {Rafiei-Ravandi}, M. and
         {Rahman}, M. and {Ransom}, S.~M. and {Renard}, A. and {Sanghavi}, P. and
         {Scholz}, P. and {Shaw}, J.~R. and {Shin}, K. and {Siegel}, S.~R. and
         {Singh}, S. and {Smegal}, R.~J. and {Smith}, K.~M. and {Stairs}, I.~H. and
         {Tendulkar}, S.~P. and {Tretyakov}, I. and {Vanderlinde}, K. and
         {Wang}, H. and {Wang}, X. and {Wulf}, D. and {Yadav}, P. and
         {Zwaniga}, A.~V.},
        title = "{Periodic activity from a fast radio burst source}",
      journal = {\nat},
         year = 2020,
        month = jun,
       volume = {582},
       number = {7812},
        pages = {351-355},
          doi = {10.1038/s41586-020-2398-2},
archivePrefix = {arXiv},
       eprint = {2001.10275},
 primaryClass = {astro-ph.HE},
       adsurl = {https://ui.adsabs.harvard.edu/abs/2020Natur.582..351C}
}

@ARTICLE{Rajwade2020,
       author = {{Rajwade}, K.~M. and {Mickaliger}, M.~B. and {Stappers}, B.~W. and
         {Morello}, V. and {Agarwal}, D. and {Bassa}, C.~G. and {Breton}, R.~P. and
         {Caleb}, M. and {Karastergiou}, A. and {Keane}, E.~F. and
         {Lorimer}, D.~R.},
        title = "{Possible periodic activity in the repeating FRB 121102}",
      journal = {\mnras},
         year = 2020,
        month = may,
       volume = {495},
       number = {4},
        pages = {3551-3558},
          doi = {10.1093/mnras/staa1237},
archivePrefix = {arXiv},
       eprint = {2003.03596},
 primaryClass = {astro-ph.HE},
       adsurl = {https://ui.adsabs.harvard.edu/abs/2020MNRAS.495.3551R}
}

@ARTICLE{Schnitzeler2012,
       author = {{Schnitzeler}, D.~H.~F.~M.},
        title = "{Modelling the Galactic distribution of free electrons}",
      journal = {\mnras},
         year = 2012,
        month = nov,
       volume = {427},
       number = {1},
        pages = {664-678},
          doi = {10.1111/j.1365-2966.2012.21869.x},
archivePrefix = {arXiv},
       eprint = {1208.3045},
 primaryClass = {astro-ph.GA},
       adsurl = {https://ui.adsabs.harvard.edu/abs/2012MNRAS.427..664S}
}

@ARTICLE{James2022a,
       author = {{James}, C.~W. and {Prochaska}, J.~X. and {Macquart}, J.~-P. and {North-Hickey}, F.~O. and {Bannister}, K.~W. and {Dunning}, A.},
        title = "{The fast radio burst population evolves, consistent with the star formation rate}",
      journal = {\mnras},
         year = 2022,
        month = feb,
       volume = {510},
       number = {1},
        pages = {L18-L23},
          doi = {10.1093/mnrasl/slab117},
archivePrefix = {arXiv},
       eprint = {2101.07998},
 primaryClass = {astro-ph.HE},
       adsurl = {https://ui.adsabs.harvard.edu/abs/2022MNRAS.510L..18J}
}

@ARTICLE{Rajwade2022,
       author = {{Rajwade}, K.~M. and {Bezuidenhout}, M.~C. and {Caleb}, M. and {Driessen}, L.~N. and {Jankowski}, F. and {Malenta}, M. and {Morello}, V. and {Sanidas}, S. and {Stappers}, B.~W. and {Surnis}, M.~P. and {Barr}, E.~D. and {Chen}, W. and {Kramer}, M. and {Wu}, J. and {Buchner}, S. and {Serylak}, M. and {Combes}, F. and {Fong}, W. and {Gupta}, N. and {Jagannathan}, P. and {Kilpatrick}, C.~D. and {Krogager}, J. -K. and {Noterdaeme}, P. and {N{\'u}nẽz}, C. and {Prochaska}, J. Xavier and {Srianand}, R. and {Tejos}, N.},
        title = "{First discoveries and localizations of Fast Radio Bursts with MeerTRAP: real-time, commensal MeerKAT survey}",
      journal = {\mnras},
         year = 2022,
        month = aug,
       volume = {514},
       number = {2},
        pages = {1961-1974},
          doi = {10.1093/mnras/stac1450},
archivePrefix = {arXiv},
       eprint = {2205.14600},
 primaryClass = {astro-ph.HE},
       adsurl = {https://ui.adsabs.harvard.edu/abs/2022MNRAS.514.1961R}
}

@ARTICLE{CHIME2021,
       author = {{CHIME/FRB Collaboration} and {Amiri}, Mandana and {Andersen}, Bridget C. and {Bandura}, Kevin and {Berger}, Sabrina and {Bhardwaj}, Mohit and {Boyce}, Michelle M. and {Boyle}, P.~J. and {Brar}, Charanjot and {Breitman}, Daniela and {Cassanelli}, Tomas and {Chawla}, Pragya and {Chen}, Tianyue and {Cliche}, J.-F. and {Cook}, Amanda and {Cubranic}, Davor and {Curtin}, Alice P. and {Deng}, Meiling and {Dobbs}, Matt and {Dong}, Fengqiu Adam and {Eadie}, Gwendolyn and {Fandino}, Mateus and {Fonseca}, Emmanuel and {Gaensler}, B.~M. and {Giri}, Utkarsh and {Good}, Deborah C. and {Halpern}, Mark and {Hill}, Alex S. and {Hinshaw}, Gary and {Josephy}, Alexander and {Kaczmarek}, Jane F. and {Kader}, Zarif and {Kania}, Joseph W. and {Kaspi}, Victoria M. and {Landecker}, T.~L. and {Lang}, Dustin and {Leung}, Calvin and {Li}, Dongzi and {Lin}, Hsiu-Hsien and {Masui}, Kiyoshi W. and {McKinven}, Ryan and {Mena-Parra}, Juan and {Merryfield}, Marcus and {Meyers}, Bradley W. and {Michilli}, Daniele and {Milutinovic}, Nikola and {Mirhosseini}, Arash and {M{\"u}nchmeyer}, Moritz and {Naidu}, Arun and {Newburgh}, Laura and {Ng}, Cherry and {Patel}, Chitrang and {Pen}, Ue-Li and {Petroff}, Emily and {Pinsonneault-Marotte}, Tristan and {Pleunis}, Ziggy and {Rafiei-Ravandi}, Masoud and {Rahman}, Mubdi and {Ransom}, Scott M. and {Renard}, Andre and {Sanghavi}, Pranav and {Scholz}, Paul and {Shaw}, J. Richard and {Shin}, Kaitlyn and {Siegel}, Seth R. and {Sikora}, Andrew E. and {Singh}, Saurabh and {Smith}, Kendrick M. and {Stairs}, Ingrid and {Tan}, Chia Min and {Tendulkar}, S.~P. and {Vanderlinde}, Keith and {Wang}, Haochen and {Wulf}, Dallas and {Zwaniga}, A.~V.},
        title = "{The First CHIME/FRB Fast Radio Burst Catalog}",
      journal = {\apjs},
         year = 2021,
        month = dec,
       volume = {257},
       number = {2},
          eid = {59},
        pages = {59},
          doi = {10.3847/1538-4365/ac33ab},
archivePrefix = {arXiv},
       eprint = {2106.04352},
 primaryClass = {astro-ph.HE},
       adsurl = {https://ui.adsabs.harvard.edu/abs/2021ApJS..257...59C}
}

@ARTICLE{Planck2018,
       author = {{Planck Collaboration} and {Aghanim}, N. and {Akrami}, Y. and {Ashdown}, M. and {Aumont}, J. and {Baccigalupi}, C. and {Ballardini}, M. and {Banday}, A.~J. and {Barreiro}, R.~B. and {Bartolo}, N. and {Basak}, S. and {Battye}, R. and {Benabed}, K. and {Bernard}, J.~-P. and {Bersanelli}, M. and {Bielewicz}, P. and {Bock}, J.~J. and {Bond}, J.~R. and {Borrill}, J. and {Bouchet}, F.~R. and {Boulanger}, F. and {Bucher}, M. and {Burigana}, C. and {Butler}, R.~C. and {Calabrese}, E. and {Cardoso}, J. -F. and {Carron}, J. and {Challinor}, A. and {Chiang}, H.~C. and {Chluba}, J. and {Colombo}, L.~P.~L. and {Combet}, C. and {Contreras}, D. and {Crill}, B.~P. and {Cuttaia}, F. and {de Bernardis}, P. and {de Zotti}, G. and {Delabrouille}, J. and {Delouis}, J. -M. and {Di Valentino}, E. and {Diego}, J.~M. and {Dor{\'e}}, O. and {Douspis}, M. and {Ducout}, A. and {Dupac}, X. and {Dusini}, S. and {Efstathiou}, G. and {Elsner}, F. and {En{\ss}lin}, T.~A. and {Eriksen}, H.~K. and {Fantaye}, Y. and {Farhang}, M. and {Fergusson}, J. and {Fernandez-Cobos}, R. and {Finelli}, F. and {Forastieri}, F. and {Frailis}, M. and {Fraisse}, A.~A. and {Franceschi}, E. and {Frolov}, A. and {Galeotta}, S. and {Galli}, S. and {Ganga}, K. and {G{\'e}nova-Santos}, R.~T. and {Gerbino}, M. and {Ghosh}, T. and {Gonz{\'a}lez-Nuevo}, J. and {G{\'o}rski}, K.~M. and {Gratton}, S. and {Gruppuso}, A. and {Gudmundsson}, J.~E. and {Hamann}, J. and {Handley}, W. and {Hansen}, F.~K. and {Herranz}, D. and {Hildebrandt}, S.~R. and {Hivon}, E. and {Huang}, Z. and {Jaffe}, A.~H. and {Jones}, W.~C. and {Karakci}, A. and {Keih{\"a}nen}, E. and {Keskitalo}, R. and {Kiiveri}, K. and {Kim}, J. and {Kisner}, T.~S. and {Knox}, L. and {Krachmalnicoff}, N. and {Kunz}, M. and {Kurki-Suonio}, H. and {Lagache}, G. and {Lamarre}, J. -M. and {Lasenby}, A. and {Lattanzi}, M. and {Lawrence}, C.~R. and {Le Jeune}, M. and {Lemos}, P. and {Lesgourgues}, J. and {Levrier}, F. and {Lewis}, A. and {Liguori}, M. and {Lilje}, P.~B. and {Lilley}, M. and {Lindholm}, V. and {L{\'o}pez-Caniego}, M. and {Lubin}, P.~M. and {Ma}, Y. -Z. and {Mac{\'\i}as-P{\'e}rez}, J.~F. and {Maggio}, G. and {Maino}, D. and {Mandolesi}, N. and {Mangilli}, A. and {Marcos-Caballero}, A. and {Maris}, M. and {Martin}, P.~G. and {Martinelli}, M. and {Mart{\'\i}nez-Gonz{\'a}lez}, E. and {Matarrese}, S. and {Mauri}, N. and {McEwen}, J.~D. and {Meinhold}, P.~R. and {Melchiorri}, A. and {Mennella}, A. and {Migliaccio}, M. and {Millea}, M. and {Mitra}, S. and {Miville-Desch{\^e}nes}, M. -A. and {Molinari}, D. and {Montier}, L. and {Morgante}, G. and {Moss}, A. and {Natoli}, P. and {N{\o}rgaard-Nielsen}, H.~U. and {Pagano}, L. and {Paoletti}, D. and {Partridge}, B. and {Patanchon}, G. and {Peiris}, H.~V. and {Perrotta}, F. and {Pettorino}, V. and {Piacentini}, F. and {Polastri}, L. and {Polenta}, G. and {Puget}, J. -L. and {Rachen}, J.~-P. and {Reinecke}, M. and {Remazeilles}, M. and {Renzi}, A. and {Rocha}, G. and {Rosset}, C. and {Roudier}, G. and {Rubi{\~n}o-Mart{\'\i}n}, J.~A. and {Ruiz-Granados}, B. and {Salvati}, L. and {Sandri}, M. and {Savelainen}, M. and {Scott}, D. and {Shellard}, E.~P.~S. and {Sirignano}, C. and {Sirri}, G. and {Spencer}, L.~D. and {Sunyaev}, R. and {Suur-Uski}, A. -S. and {Tauber}, J.~A. and {Tavagnacco}, D. and {Tenti}, M. and {Toffolatti}, L. and {Tomasi}, M. and {Trombetti}, T. and {Valenziano}, L. and {Valiviita}, J. and {Van Tent}, B. and {Vibert}, L. and {Vielva}, P. and {Villa}, F. and {Vittorio}, N. and {Wandelt}, B.~D. and {Wehus}, I.~K. and {White}, M. and {White}, S.~D.~M. and {Zacchei}, A. and {Zonca}, A.},
        title = "{Planck 2018 results. VI. Cosmological parameters}",
      journal = {\aap},
         year = 2020,
        month = sep,
       volume = {641},
          eid = {A6},
        pages = {A6},
          doi = {10.1051/0004-6361/201833910},
archivePrefix = {arXiv},
       eprint = {1807.06209},
 primaryClass = {astro-ph.CO},
       adsurl = {https://ui.adsabs.harvard.edu/abs/2020A&A...641A...6P}
}

@ARTICLE{Li2021,
       author = {{Li}, D. and {Wang}, P. and {Zhu}, W.~W. and {Zhang}, B. and {Zhang}, X.~X. and {Duan}, R. and {Zhang}, Y.~K. and {Feng}, Y. and {Tang}, N.~Y. and {Chatterjee}, S. and {Cordes}, J.~M. and {Cruces}, M. and {Dai}, S. and {Gajjar}, V. and {Hobbs}, G. and {Jin}, C. and {Kramer}, M. and {Lorimer}, D.~R. and {Miao}, C.~C. and {Niu}, C.~H. and {Niu}, J.~R. and {Pan}, Z.~C. and {Qian}, L. and {Spitler}, L. and {Werthimer}, D. and {Zhang}, G.~Q. and {Wang}, F.~Y. and {Xie}, X.~Y. and {Yue}, Y.~L. and {Zhang}, L. and {Zhi}, Q.~J. and {Zhu}, Y.},
        title = "{A bimodal burst energy distribution of a repeating fast radio burst source}",
      journal = {\nat},
         year = 2021,
        month = oct,
       volume = {598},
       number = {7880},
        pages = {267-271},
          doi = {10.1038/s41586-021-03878-5},
archivePrefix = {arXiv},
       eprint = {2107.08205},
 primaryClass = {astro-ph.HE},
       adsurl = {https://ui.adsabs.harvard.edu/abs/2021Natur.598..267L}
}

@ARTICLE{Zhang2023,
       author = {{Zhang}, Bing},
        title = "{The physics of fast radio bursts}",
      journal = {Reviews of Modern Physics},
         year = 2023,
        month = jul,
       volume = {95},
       number = {3},
          eid = {035005},
        pages = {035005},
          doi = {10.1103/RevModPhys.95.035005},
archivePrefix = {arXiv},
       eprint = {2212.03972},
 primaryClass = {astro-ph.HE},
       adsurl = {https://ui.adsabs.harvard.edu/abs/2023RvMP...95c5005Z}
}

@ARTICLE{SH0ES2021,
       author = {{Riess}, Adam G. and {Yuan}, Wenlong and {Macri}, Lucas M. and {Scolnic}, Dan and {Brout}, Dillon and {Casertano}, Stefano and {Jones}, David O. and {Murakami}, Yukei and {Anand}, Gagandeep S. and {Breuval}, Louise and {Brink}, Thomas G. and {Filippenko}, Alexei V. and {Hoffmann}, Samantha and {Jha}, Saurabh W. and {D'arcy Kenworthy}, W. and {Mackenty}, John and {Stahl}, Benjamin E. and {Zheng}, WeiKang},
        title = "{A Comprehensive Measurement of the Local Value of the Hubble Constant with 1 km s$^{-1}$ Mpc$^{-1}$ Uncertainty from the Hubble Space Telescope and the SH0ES Team}",
      journal = {\apjl},
         year = 2022,
        month = jul,
       volume = {934},
       number = {1},
          eid = {L7},
        pages = {L7},
          doi = {10.3847/2041-8213/ac5c5b},
archivePrefix = {arXiv},
       eprint = {2112.04510},
 primaryClass = {astro-ph.CO},
       adsurl = {https://ui.adsabs.harvard.edu/abs/2022ApJ...934L...7R}
}

@ARTICLE{Cordes2002,
       author = {{Cordes}, J.~M. and {Lazio}, T.~J.~W.},
        title = "{NE2001.I. A New Model for the Galactic Distribution of Free Electrons and its Fluctuations}",
      journal = {arXiv e-prints},
         year = 2002,
        month = jul,
          eid = {astro-ph/0207156},
        pages = {astro-ph/0207156},
archivePrefix = {arXiv},
       eprint = {astro-ph/0207156},
 primaryClass = {astro-ph},
       adsurl = {https://ui.adsabs.harvard.edu/abs/2002astro.ph..7156C}
}

@ARTICLE{Prochaska2019a,
       author = {{Prochaska}, J. Xavier and {Zheng}, Yong},
        title = "{Probing Galactic haloes with fast radio bursts}",
      journal = {\mnras},
         year = {2019},
        month = may,
       volume = {485},
       number = {1},
        pages = {648-665},
          doi = {10.1093/mnras/stz261},
archivePrefix = {arXiv},
       eprint = {1901.11051},
 primaryClass = {astro-ph.GA},
       adsurl = {https://ui.adsabs.harvard.edu/abs/2019MNRAS.485..648P}
}

@ARTICLE{Hoffmann2024,
       author = {{Hoffmann}, J. and {James}, C.~W. and {Qiu}, H. and {Glowacki}, M. and {Bannister}, K.~W. and {Gupta}, V. and {Prochaska}, J.~X. and {Bera}, A. and {Deller}, A.~T. and {Gourdji}, K. and {Marnoch}, L. and {Ryder}, S.~D. and {Scott}, D.~R. and {Shannon}, R.~M. and {Tejos}, N.},
        title = "{The impact of the FREDDA dedispersion algorithm on H$_{0}$ estimations with fast radio bursts}",
      journal = {\mnras},
         year = 2024,
        month = feb,
       volume = {528},
       number = {2},
        pages = {1583-1595},
          doi = {10.1093/mnras/stae131},
       adsurl = {https://ui.adsabs.harvard.edu/abs/2024MNRAS.528.1583H}
}

@ARTICLE{Cook2026_cat2repeaters,
       author = {{Cook}, Amanda M. and {Shin}, Kaitlyn and {Pleunis}, Ziggy and {Fine}, Maxwell and {Jain}, Naman and {Bingham}, Derek and {Curtin}, Alice P. and {Eadie}, Gwendolyn and {Gaensler}, B.~M. and {Hessels}, Jason W.~T. and {Leung}, Calvin and {Main}, Robert and {Mulyk}, Nicole and {Pandhi}, Ayush and {Scholz}, Paul and {Siegel}, Seth R. and {Stenning}, David C. and {Abbott}, Thomas C. and {Andersen}, Bridget C. and {Bhardwaj}, Mohit and {Cai}, Alice and {Chatterjee}, Shami and {Dong}, Fengqiu Adam and {Fonseca}, Emmanuel and {Hewitt}, Dant{\'e} M. and {Joseph}, Ronniy C. and {Kahinga}, Lordrick and {Lazda}, Mattias and {Kaspi}, Victoria M. and {Khan}, Afrokk and {Kharel}, Bikash and {Mas-Ribas}, Lluis and {Masui}, Kiyoshi W. and {McGregor}, Kyle and {Michilli}, Daniele and {Mckinven}, Ryan and {Ng}, Mason and {Nimmo}, Kenzie and {Shivraj Patil}, Swarali and {Pearlman}, Aaron B. and {Sammons}, Mawson W. and {Sand}, Ketan R. and {Sedaei Oghani}, Aylar and {Shah}, Vishwangi and {Smith}, Kendrick and {Stairs}, Ingrid and {Zegmott}, Tarik J.},
        title = "{Discovery of 30 Repeating Fast Radio Burst Sources and Uniform Population Statistics of 80 Repeating Sources from CHIME/FRB}",
      journal = {arXiv e-prints},
         year = 2026,
        month = may,
          eid = {arXiv:2605.08410},
        pages = {arXiv:2605.08410},
          doi = {10.48550/arXiv.2605.08410},
archivePrefix = {arXiv},
       eprint = {2605.08410},
 primaryClass = {astro-ph.HE},
       adsurl = {https://ui.adsabs.harvard.edu/abs/2026arXiv260508410C}
}

@ARTICLE{Caleb2025z2,
       author = {{Caleb}, Manisha and {Nanayakkara}, Themiya and {Stappers}, Benjamin and {Pastor-Marazuela}, In{\'e}s and {Khrykin}, Ilya S. and {Glazebrook}, Karl and {Tejos}, Nicolas and {Prochaska}, J. Xavier and {Rajwade}, Kaustubh and {Mas-Ribas}, Lluis and {Driessen}, Laura N. and {Fong}, Wen-fai and {Gordon}, Alexa C. and {Hoffmann}, Jordan and {James}, Clancy W. and {Jankowski}, Fabian and {Kahinga}, Lordrick and {Kramer}, Michael and {Simha}, Sunil and {Barr}, Ewan D. and {Christiaan Bezuidenhout}, Mechiel and {Deng}, Xihan and {Lin}, Zeren and {Marnoch}, Lachlan and {Martin}, Christopher D. and {Nugent}, Anya and {Shaji}, Kavya and {Tian}, Jun},
        title = "{A fast radio burst from the first 3 billion years of the Universe}",
      journal = {arXiv e-prints},
         year = 2025,
        month = aug,
          eid = {arXiv:2508.01648},
        pages = {arXiv:2508.01648},
          doi = {10.48550/arXiv.2508.01648},
archivePrefix = {arXiv},
       eprint = {2508.01648},
 primaryClass = {astro-ph.HE},
       adsurl = {https://ui.adsabs.harvard.edu/abs/2025arXiv250801648C}
}

@ARTICLE{Jankowski2023,
       author = {{Jankowski}, F. and {Bezuidenhout}, M.~C. and {Caleb}, M. and {Driessen}, L.~N. and {Malenta}, M. and {Morello}, V. and {Rajwade}, K.~M. and {Sanidas}, S. and {Stappers}, B.~W. and {Surnis}, M.~P. and {Barr}, E.~D. and {Chen}, W. and {Kramer}, M. and {Wu}, J. and {Buchner}, S. and {Serylak}, M. and {Prochaska}, J. Xavier},
        title = "{A sample of fast radio bursts discovered and localized with MeerTRAP at the MeerKAT telescope}",
      journal = {\mnras},
         year = 2023,
        month = sep,
       volume = {524},
       number = {3},
        pages = {4275-4295},
          doi = {10.1093/mnras/stad2041},
archivePrefix = {arXiv},
       eprint = {2302.10107},
 primaryClass = {astro-ph.HE},
       adsurl = {https://ui.adsabs.harvard.edu/abs/2023MNRAS.524.4275J}
}

@ARTICLE{Sharma2026,
       author = {{Sharma}, Kritti and {Krause}, Elisabeth and {Ravi}, Vikram and {Anbajagane}, Dhayaa and {Connor}, Liam and {Kimmy Wu}, W.~L. and {Ferraro}, Simone and {Grandis}, Sebastian and {Alonso}, David and {Chiang}, Yi-Kuan and {Law}, Casey J. and {Pranjal R.}, S. and {McCarty}, Samuel and {Pandey}, Shivam},
        title = "{Backlighting the Cosmic Web with Fast Radio Bursts: An Anthology of Dispersion Measure Cross-Correlations with Large-Scale Structure and Baryon Tracers}",
      journal = {arXiv e-prints},
         year = 2026,
        month = apr,
          eid = {arXiv:2604.22105},
        pages = {arXiv:2604.22105},
          doi = {10.48550/arXiv.2604.22105},
archivePrefix = {arXiv},
       eprint = {2604.22105},
 primaryClass = {astro-ph.CO},
       adsurl = {https://ui.adsabs.harvard.edu/abs/2026arXiv260422105S}
}

@ARTICLE{StaveleySmith1996,
       author = {{Staveley-Smith}, L. and {Wilson}, W.~E. and {Bird}, T.~S. and {Disney}, M.~J. and {Ekers}, R.~D. and {Freeman}, K.~C. and {Haynes}, R.~F. and {Sinclair}, M.~W. and {Vaile}, R.~A. and {Webster}, R.~L. and {Wright}, A.~E.},
        title = "{The Parkes 21 CM multibeam receiver}",
      journal = {\pasa},
         year = 1996,
        month = nov,
       volume = {13},
       number = {3},
        pages = {243-248},
          doi = {10.1017/S1323358000020919},
       adsurl = {https://ui.adsabs.harvard.edu/abs/1996PASA...13..243S}
}

@ARTICLE{Taylor2014,
       author = {{Taylor}, Matt and {Cinabro}, David and {Dilday}, Ben and {Galbany}, Lluis and {Gupta}, Ravi R. and {Kessler}, R. and {Marriner}, John and {Nichol}, Robert C. and {Richmond}, Michael and {Schneider}, Donald P. and {Sollerman}, Jesper},
        title = "{The Core Collapse Supernova Rate from the SDSS-II Supernova Survey}",
      journal = {\apj},
         year = 2014,
        month = sep,
       volume = {792},
       number = {2},
          eid = {135},
        pages = {135},
          doi = {10.1088/0004-637X/792/2/135},
archivePrefix = {arXiv},
       eprint = {1407.0999},
 primaryClass = {astro-ph.SR},
       adsurl = {https://ui.adsabs.harvard.edu/abs/2014ApJ...792..135T}
}

@ARTICLE{Sherman2023,
       author = {{Sherman}, Myles B. and {Connor}, Liam and {Ravi}, Vikram and {Law}, Casey and {Chen}, Ge and {Catha}, Morgan and {Faber}, Jakob T. and {Hallinan}, Gregg and {Harnach}, Charlie and {Hellbourg}, Greg and {Hobbs}, Rick and {Hodge}, David and {Hodges}, Mark and {Lamb}, James W. and {Rasmussen}, Paul and {Sharma}, Kritti and {Shi}, Jun and {Simard}, Dana and {Somalwar}, Jean and {Squillace}, Reynier and {Weinreb}, Sander and {Woody}, David P. and {Yadlapalli}, Nitika},
        title = "{Deep Synoptic Array Science: Polarimetry of 25 New Fast Radio Bursts Provides Insights into their Origins}",
      journal = {arXiv e-prints},
         year = 2023,
        month = aug,
          eid = {arXiv:2308.06813},
        pages = {arXiv:2308.06813},
          doi = {10.48550/arXiv.2308.06813},
archivePrefix = {arXiv},
       eprint = {2308.06813},
 primaryClass = {astro-ph.HE},
       adsurl = {https://ui.adsabs.harvard.edu/abs/2023arXiv230806813S}
}

@ARTICLE{Colpi2000,
       author = {{Colpi}, Monica and {Geppert}, Ulrich and {Page}, Dany},
        title = "{Period Clustering of the Anomalous X-Ray Pulsars and Magnetic Field Decay in Magnetars}",
      journal = {\apjl},
         year = 2000,
        month = jan,
       volume = {529},
       number = {1},
        pages = {L29-L32},
          doi = {10.1086/312448},
archivePrefix = {arXiv},
       eprint = {astro-ph/9912066},
 primaryClass = {astro-ph},
       adsurl = {https://ui.adsabs.harvard.edu/abs/2000ApJ...529L..29C}
}

@ARTICLE{Cruces2021,
       author = {{Cruces}, M. and {Spitler}, L.~G. and {Scholz}, P. and {Lynch}, R. and {Seymour}, A. and {Hessels}, J.~W.~T. and {Gouiff{\'e}s}, C. and {Hilmarsson}, G.~H. and {Kramer}, M. and {Munjal}, S.},
        title = "{Repeating behaviour of FRB 121102: periodicity, waiting times, and energy distribution}",
      journal = {\mnras},
         year = 2021,
        month = jan,
       volume = {500},
       number = {1},
        pages = {448-463},
          doi = {10.1093/mnras/staa3223},
archivePrefix = {arXiv},
       eprint = {2008.03461},
 primaryClass = {astro-ph.HE},
       adsurl = {https://ui.adsabs.harvard.edu/abs/2021MNRAS.500..448C}
}

@ARTICLE{Braga2025,
       author = {{Braga}, C.~A. and {Cruces}, M. and {Cassanelli}, T. and {Espinoza-Dupouy}, M.~C. and {Rodriguez}, L. and {Spitler}, L.~G. and {Vera-Casanova}, J. and {Limaye}, P.},
        title = "{FRB 20121102A monitoring: Updated periodicity in the L band}",
      journal = {\aap},
         year = 2025,
        month = jan,
       volume = {693},
          eid = {A40},
        pages = {A40},
          doi = {10.1051/0004-6361/202451905},
archivePrefix = {arXiv},
       eprint = {2408.12567},
 primaryClass = {astro-ph.HE},
       adsurl = {https://ui.adsabs.harvard.edu/abs/2025A&A...693A..40B}
}

@ARTICLE{Macquart2019,
       author = {{Macquart}, J.-P. and {Shannon}, R.~M. and {Bannister}, K.~W. and {James}, C.~W. and {Ekers}, R.~D. and {Bunton}, J.~D.},
        title = "{The Spectral Properties of the Bright Fast Radio Burst Population}",
      journal = {\apjl},
         year = 2019,
        month = feb,
       volume = {872},
       number = {2},
          eid = {L19},
        pages = {L19},
          doi = {10.3847/2041-8213/ab03d6},
archivePrefix = {arXiv},
       eprint = {1810.04353},
 primaryClass = {astro-ph.HE},
       adsurl = {https://ui.adsabs.harvard.edu/abs/2019ApJ...872L..19M}
}

@ARTICLE{james2022b,
       author = {{James}, C.~W. and {Ghosh}, E.~M. and {Prochaska}, J.~X. and {Bannister}, K.~W. and {Bhandari}, S. and {Day}, C.~K. and {Deller}, A.~T. and {Glowacki}, M. and {Gordon}, A.~C. and {Heintz}, K.~E. and {Marnoch}, L. and {Ryder}, S.~D. and {Scott}, D.~R. and {Shannon}, R.~M. and {Tejos}, N.},
        title = "{A measurement of Hubble's Constant using Fast Radio Bursts}",
      journal = {\mnras},
         year = {2022},
        month = nov,
       volume = {516},
       number = {4},
        pages = {4862-4881},
          doi = {10.1093/mnras/stac2524},
archivePrefix = {arXiv},
       eprint = {2208.00819},
 primaryClass = {astro-ph.CO},
       adsurl = {https://ui.adsabs.harvard.edu/abs/2022MNRAS.516.4862J}
}

@ARTICLE{Gardenier2019_FRBpoppy,
       author = {{Gardenier}, D.~W. and {van Leeuwen}, J. and {Connor}, L. and {Petroff}, E.},
        title = "{Synthesising the intrinsic FRB population using frbpoppy}",
      journal = {\aap},
         year = 2019,
        month = dec,
       volume = {632},
          eid = {A125},
        pages = {A125},
          doi = {10.1051/0004-6361/201936404},
       adsurl = {https://ui.adsabs.harvard.edu/abs/2019A&A...632A.125G}
}

@ARTICLE{Macquart2010,
       author = {{Macquart}, Jean-Pierre and {Bailes}, M. and {Bhat}, N.~D.~R. and {Bower}, G.~C. and {Bunton}, J.~D. and {Chatterjee}, S. and {Colegate}, T. and {Cordes}, J.~M. and {D'Addario}, L. and {Deller}, A. and {Dodson}, R. and {Fender}, R. and {Haines}, K. and {Halll}, P. and {Harris}, C. and {Hotan}, A. and {Johnston}, S. and {Jones}, D.~L. and {Keith}, M. and {Koay}, J.~Y. and {Lazio}, T.~J.~W. and {Majid}, W. and {Murphy}, T. and {Navarro}, R. and {Phillips}, C. and {Quinn}, P. and {Preston}, R.~A. and {Stansby}, B. and {Stairs}, I. and {Stappers}, B. and {Staveley-Smith}, L. and {Tingay}, S. and {Thompson}, D. and {van Straten}, W. and {Wagstaff}, K. and {Warren}, M. and {Wayth}, R. and {Wen}, L. and {CRAFT Collaboration}},
        title = "{The Commensal Real-Time ASKAP Fast-Transients (CRAFT) Survey}",
      journal = {\pasa},
         year = 2010,
        month = jun,
       volume = {27},
       number = {3},
        pages = {272-282},
          doi = {10.1071/AS09082},
archivePrefix = {arXiv},
       eprint = {1001.2958},
 primaryClass = {astro-ph.HE},
       adsurl = {https://ui.adsabs.harvard.edu/abs/2010PASA...27..272M}
}

@INPROCEEDINGS{Ravi2021,
       author = {{Ravi}, V. and {Hallinan}, G. and {Deep Synoptic Array Team}},
        title = "{The Deep Synoptic Array}",
    booktitle = {American Astronomical Society Meeting Abstracts},
         year = 2021,
       series = {American Astronomical Society Meeting Abstracts},
       volume = {53},
        month = jan,
          eid = {316.04},
        pages = {316.04},
       adsurl = {https://ui.adsabs.harvard.edu/abs/2021AAS...23731604R}
}

@INPROCEEDINGS{MeerTRAP2018,
       author = {{Sanidas}, S. and {Caleb}, M. and {Driessen}, L. and {Morello}, V. and {Rajwade}, K. and {Stappers}, B.~W.},
        title = "{MeerTRAP: A pulsar and fast transients survey with MeerKAT}",
    booktitle = {Pulsar Astrophysics the Next Fifty Years},
         year = 2018,
       editor = {{Weltevrede}, P. and {Perera}, B.~B.~P. and {Preston}, L.~L. and {Sanidas}, S.},
       volume = {337},
        month = aug,
        pages = {406-407},
          doi = {10.1017/S1743921317009310},
       adsurl = {https://ui.adsabs.harvard.edu/abs/2018IAUS..337..406S}
}

@ARTICLE{Khrykin2024,
       author = {{Khrykin}, Ilya S. and {Ata}, Metin and {Lee}, Khee-Gan and {Simha}, Sunil and {Huang}, Yuxin and {Prochaska}, J. Xavier and {Tejos}, Nicolas and {Bannister}, Keith W. and {Cooke}, Jeff and {Day}, Cherie K. and {Deller}, Adam and {Glowacki}, Marcin and {Gordon}, Alexa C. and {James}, Clancy W. and {Marnoch}, Lachlan and {Shannon}, Ryan. M. and {Zhang}, Jielai and {Bernales-Cortes}, Lucas},
        title = "{FLIMFLAM DR1: The First Constraints on the Cosmic Baryon Distribution from Eight Fast Radio Burst Sight Lines}",
      journal = {\apj},
         year = 2024,
        month = oct,
       volume = {973},
       number = {2},
          eid = {151},
        pages = {151},
          doi = {10.3847/1538-4357/ad6567},
archivePrefix = {arXiv},
       eprint = {2402.00505},
 primaryClass = {astro-ph.GA},
       adsurl = {https://ui.adsabs.harvard.edu/abs/2024ApJ...973..151K}
}

@misc{zdm,
  author = {{James}, C. W. and {Prochaska}, J. X. and {Ghosh}, E. M.},
  title = {zdm},
  publisher = {GitHub},
  journal = {GitHub repository},
  url = {https://zenodo.org/record/5213780#.YRxh5BMzZKA},
  version = {0.1},
  date = {2021-08-18},
  year = 2021
  }

@ARTICLE{Spitler2016,
       author = {{Spitler}, L.~G. and {Scholz}, P. and {Hessels}, J.~W.~T. and {Bogdanov}, S. and {Brazier}, A. and {Camilo}, F. and {Chatterjee}, S. and {Cordes}, J.~M. and {Crawford}, F. and {Deneva}, J. and {Ferdman}, R.~D. and {Freire}, P.~C.~C. and {Kaspi}, V.~M. and {Lazarus}, P. and {Lynch}, R. and {Madsen}, E.~C. and {McLaughlin}, M.~A. and {Patel}, C. and {Ransom}, S.~M. and {Seymour}, A. and {Stairs}, I.~H. and {Stappers}, B.~W. and {van Leeuwen}, J. and {Zhu}, W.~W.},
        title = "{A repeating fast radio burst}",
      journal = {\nat},
         year = 2016,
        month = mar,
       volume = {531},
       number = {7593},
        pages = {202-205},
          doi = {10.1038/nature17168},
archivePrefix = {arXiv},
       eprint = {1603.00581},
 primaryClass = {astro-ph.HE},
       adsurl = {https://ui.adsabs.harvard.edu/abs/2016Natur.531..202S}
}

@ARTICLE{Hoffmann2026,
       author = {{Hoffmann}, Jordan Luke and {James}, Clancy and {Prochaska}, Xavier and {Glowacki}, Marcin},
        title = "{I can see your halo: Constraining the MilkyWay halo DM with FRB population studies}",
      journal = {\pasa},
         year = 2026,
        month = jan,
       volume = {43},
          eid = {e017},
        pages = {e017},
          doi = {10.1017/pasa.2026.10143},
archivePrefix = {arXiv},
       eprint = {2601.05496},
 primaryClass = {astro-ph.GA},
       adsurl = {https://ui.adsabs.harvard.edu/abs/2026PASA...43...17H}
}

@ARTICLE{Scott2025HTR,
       author = {{Scott}, Danica R. and {Dial}, Tyson and {Bera}, Apurba and {Deller}, Adam T. and {Glowacki}, Marcin and {Gourdji}, Kelly and {James}, Clancy W. and {Shannon}, Ryan M. and {Bannister}, Keith W. and {Ekers}, Ron D. and {Paterson}, Jasper and {Sammons}, Mawson and {Sutinjo}, Adrian T. and {Uttarkar}, Pavan A.},
        title = "{High-time-resolution properties of 35 fast radio bursts detected by the Commensal Real-time ASKAP Fast Transients survey}",
      journal = {\pasa},
         year = 2025,
        month = oct,
       volume = {42},
          eid = {e133},
        pages = {e133},
          doi = {10.1017/pasa.2025.10103},
archivePrefix = {arXiv},
       eprint = {2505.17497},
 primaryClass = {astro-ph.HE},
       adsurl = {https://ui.adsabs.harvard.edu/abs/2025PASA...42..133S}
}

@ARTICLE{CHIME2026cat2,
       author = {{FRB Collaboration} and {Abbott}, Thomas and {Andersen}, Bridget C. and {Andrew}, Shion and {Bandura}, Kevin and {Bhardwaj}, Mohit and {Bhusare}, Yash and {Brar}, Charanjot and {Cassanelli}, Tomas and {Chatterjee}, Shami and {Cliche}, Jean-Francois and {Cook}, Amanda M. and {Curtin}, Alice and {Dobbs}, Matt and {Dong}, Fengqiu Adam and {Eadie}, Gwendolyn and {Eftekhari}, Tarraneh and {Fonseca}, Emmanuel and {Gaensler}, B.~M. and {Good}, Deborah and {Halpern}, Mark and {Hessels}, Jason W.~T. and {Ibik}, Adaeze and {Jain}, Naman and {Joseph}, Ronniy C. and {Kader}, Zarif and {Kaspi}, Victoria M. and {Khan}, Afrokk and {Kharel}, Bikash and {Kumar}, Ajay and {Landecker}, T.~L. and {Lang}, Dustin and {Lanman}, Adam E. and {L'Argent}, Magnus and {Lazda}, Mattias and {Leung}, Calvin and {Li}, Dong Zi and {Lintott}, Chris J. and {Main}, Robert and {Masui}, Kiyoshi W. and {Mate}, Sujay and {McGregor}, Kyle and {Mckinven}, Ryan and {Mena-Parra}, Juan and {Meyers}, Bradley W. and {Michilli}, Daniele and {Ng}, Cherry and {Ng}, Mason and {Nimmo}, Kenzie and {Noble}, Gavin and {Pandhi}, Ayush and {Patil}, Swarali S. and {Pearlman}, Aaron B. and {Pen}, Ue-Li and {Pleunis}, Ziggy and {Prochaska}, J. Xavier and {Rafiei-Ravandi}, Masoud and {Ransom}, Scott and {Renard}, Andre and {Sammons}, Mawson W. and {Sand}, Ketan R. and {Scholz}, Paul and {Shah}, Vishwangi and {Shin}, Kaitlyn and {Siegel}, Seth R. and {Sirota}, Sloane and {Smith}, Kendrick and {Stairs}, Ingrid and {Stenning}, David C. and {Tendulkar}, Shriharsh P. and {Vanderlinde}, Keith and {Walmsley}, Mike and {Wang}, Haochen and {Wulf}, Dallas},
        title = "{The Second CHIME/FRB Catalog of Fast Radio Bursts}",
      journal = {arXiv e-prints},
         year = 2026,
        month = jan,
          eid = {arXiv:2601.09399},
        pages = {arXiv:2601.09399},
          doi = {10.48550/arXiv.2601.09399},
archivePrefix = {arXiv},
       eprint = {2601.09399},
 primaryClass = {astro-ph.HE},
       adsurl = {https://ui.adsabs.harvard.edu/abs/2026arXiv260109399F}
}

@ARTICLE{deboer2009,
       author = {{DeBoer}, D.~R. and {Gough}, R.~G. and {Bunton}, J.~D. and {Cornwell}, T.~J. and {Beresford}, R.~J. and {Johnston}, S. and {Feain}, I.~J. and {Schinckel}, A.~E. and {Jackson}, C.~A. and {Kesteven}, M.~J. and {Chippendale}, A. and {Hampson}, G.~A. and {O'Sullivan}, J.~D. and {Hay}, S.~G. and {Jacka}, C.~E. and {Sweetnam}, T.~W. and {Storey}, M.~C. and {Ball}, L. and {Boyle}, B.~J.},
        title = "{Australian SKA Pathfinder: A High-Dynamic Range Wide-Field of View Survey Telescope}",
      journal = {IEEE Proceedings},
         year = 2009,
        month = aug,
       volume = {97},
       number = {8},
        pages = {1507-1521},
          doi = {10.1109/JPROC.2009.2016516},
       adsurl = {https://ui.adsabs.harvard.edu/abs/2009IEEEP..97.1507D}
}

@INPROCEEDINGS{jonas2016,
       author = {{Jonas}, J. and {MeerKAT Team}},
        title = "{The MeerKAT Radio Telescope}",
    booktitle = {MeerKAT Science: On the Pathway to the SKA},
         year = 2016,
        month = jan,
          eid = {1},
        pages = {1},
          doi = {10.22323/1.277.0001},
       adsurl = {https://ui.adsabs.harvard.edu/abs/2016mks..confE...1J}
}

@ARTICLE{Chawla2022,
       author = {{Chawla}, P. and {Kaspi}, V.~M. and {Ransom}, S.~M. and {Bhardwaj}, M. and {Boyle}, P.~J. and {Breitman}, D. and {Cassanelli}, T. and {Cubranic}, D. and {Dong}, F.~Q. and {Fonseca}, E. and {Gaensler}, B.~M. and {Giri}, U. and {Josephy}, A. and {Kaczmarek}, J.~F. and {Leung}, C. and {Masui}, K.~W. and {Mena-Parra}, J. and {Merryfield}, M. and {Michilli}, D. and {M{\"u}nchmeyer}, M. and {Ng}, C. and {Patel}, C. and {Pearlman}, A.~B. and {Petroff}, E. and {Pleunis}, Z. and {Rahman}, M. and {Sanghavi}, P. and {Shin}, K. and {Smith}, K.~M. and {Stairs}, I. and {Tendulkar}, S.~P.},
        title = "{Modeling Fast Radio Burst Dispersion and Scattering Properties in the First CHIME/FRB Catalog}",
      journal = {\apj},
         year = 2022,
        month = mar,
       volume = {927},
       number = {1},
          eid = {35},
        pages = {35},
          doi = {10.3847/1538-4357/ac49e1},
archivePrefix = {arXiv},
       eprint = {2107.10858},
 primaryClass = {astro-ph.HE},
       adsurl = {https://ui.adsabs.harvard.edu/abs/2022ApJ...927...35C}
}

@ARTICLE{shin2023,
       author = {{Shin}, Kaitlyn and {Masui}, Kiyoshi W. and {Bhardwaj}, Mohit and {Cassanelli}, Tomas and {Chawla}, Pragya and {Dobbs}, Matt and {Dong}, Fengqiu Adam and {Fonseca}, Emmanuel and {Gaensler}, B.~M. and {Herrera-Mart{\'\i}n}, Antonio and {Kaczmarek}, Jane and {Kaspi}, Victoria and {Leung}, Calvin and {Merryfield}, Marcus and {Michilli}, Daniele and {M{\"u}nchmeyer}, Moritz and {Pearlman}, Aaron B. and {Rafiei-Ravandi}, Masoud and {Smith}, Kendrick and {Stairs}, Ingrid and {Tendulkar}, Shriharsh P.},
        title = "{Inferring the Energy and Distance Distributions of Fast Radio Bursts Using the First CHIME/FRB Catalog}",
      journal = {\apj},
         year = 2023,
        month = feb,
       volume = {944},
       number = {1},
          eid = {105},
        pages = {105},
          doi = {10.3847/1538-4357/acaf06},
archivePrefix = {arXiv},
       eprint = {2207.14316},
 primaryClass = {astro-ph.HE},
       adsurl = {https://ui.adsabs.harvard.edu/abs/2023ApJ...944..105S}
}

@ARTICLE{Hoffmann2025,
       author = {{Hoffmann}, J. and {James}, C.~W. and {Glowacki}, M. and {Prochaska}, X. and {Gordon}, A. and {Deller}, A and {Shannon}, R.~M. and {Ryder}, S.~D.},
        title = "{Modelling DSA, FAST, and CRAFT surveys in a z-DM analysis and constraining a minimum FRB energy}",
      journal = {\pasa},
         year = 2025,
        month = jan,
       volume = {42},
          eid = {e017},
        pages = {e017},
          doi = {10.1017/pasa.2024.127},
archivePrefix = {arXiv},
       eprint = {2408.04878},
 primaryClass = {astro-ph.CO},
       adsurl = {https://ui.adsabs.harvard.edu/abs/2025PASA...42...17H}
}

@ARTICLE{PastorMarazuela2025,
       author = {{Pastor-Marazuela}, In{\'e}s and {Gordon}, Alexa C. and {Stappers}, Ben and {Khrykin}, Ilya S. and {Tejos}, Nicolas and {Rajwade}, Kaustubh and {Caleb}, Manisha and {Surnis}, Mayuresh P. and {Driessen}, Laura N. and {Simha}, Sunil and {Tian}, Jun and {Prochaska}, J. Xavier and {Barr}, Ewan and {Buchner}, Sarah and {Fong}, Wen-Fai and {Jankowski}, Fabian and {Kahinga}, Lordrick and {Kilpatrick}, Charles D. and {Kramer}, Michael and {Mas-Ribas}, Lluis and {Hennawi}, Joseph},
        title = "{Localisation and host galaxy identification of new Fast Radio Bursts with MeerKAT}",
      journal = {\mnras},
         year = 2025,
        month = dec,
          doi = {10.1093/mnras/staf2144},
archivePrefix = {arXiv},
       eprint = {2507.05982},
 primaryClass = {astro-ph.HE},
       adsurl = {https://ui.adsabs.harvard.edu/abs/2025MNRAS.tmp.2025P}
}

@ARTICLE{CHIME2025Outriggers,
       author = {{CHIME/FRB Collaboration} and {Amiri}, Mandana and {Amouyal}, Daniel and {Andersen}, Bridget C. and {Andrew}, Shion and {Bandura}, Kevin and {Bhardwaj}, Mohit and {Boyle}, P.~J. and {Brar}, Charanjot and {Cassity}, Alyssa and {Chatterjee}, Shami and {Curtin}, Alice P. and {Dobbs}, Matt and {Dong}, Fengqiu Adam and {Dong}, Yuxin and {Eadie}, Gwendolyn M. and {Eftekhari}, Tarraneh and {Fong}, Wen-Fai and {Fonseca}, Emmanuel and {Gaensler}, B.~M. and {Halpern}, Mark and {Hessels}, Jason W.~T. and {Hopkins}, Hans and {Ibik}, Adaeze L. and {Joseph}, Ronniy C. and {Kaczmarek}, Jane and {Kahinga}, Lordrick and {Kaspi}, Victoria and {Khairy}, Kholoud and {Kilpatrick}, Charles D. and {Lanman}, Adam E. and {Lazda}, Mattias and {Leung}, Calvin and {Main}, Robert and {Mas-Ribas}, Lluis and {Masui}, Kiyoshi W. and {McKinven}, Ryan and {Mena-Parra}, Juan and {Meyers}, Bradley W. and {Michilli}, Daniele and {Milutinovic}, Nikola and {Nimmo}, Kenzie and {Noble}, Gavin and {Pandhi}, Ayush and {Patil}, Swarali Shivraj and {Pearlman}, Aaron B. and {Petroff}, Emily and {Pleunis}, Ziggy and {Prochaska}, J. Xavier and {Rafiei-Ravandi}, Masoud and {Rahman}, Mubdi and {Renard}, Andre and {Sammons}, Mawson W. and {Sand}, Ketan R. and {Scholz}, Paul and {Shah}, Vishwangi and {Shin}, Kaitlyn and {Siegel}, Seth R. and {Simha}, Sunil and {Smith}, Kendrick and {Stairs}, Ingrid and {Vanderlinde}, Keith and {Wang}, Haochen and {Wulf}, Dallas and {Zegmott}, Tarik J.},
        title = "{A Catalog of Local Universe Fast Radio Bursts from CHIME/FRB and the KKO}",
      journal = {\apjs},
         year = 2025,
        month = sep,
       volume = {280},
       number = {1},
          eid = {6},
        pages = {6},
          doi = {10.3847/1538-4365/addbda},
archivePrefix = {arXiv},
       eprint = {2502.11217},
 primaryClass = {astro-ph.HE},
       adsurl = {https://ui.adsabs.harvard.edu/abs/2025ApJS..280....6C}
}

@ARTICLE{CHIMEbaseband2024,
       author = {{CHIME/FRB Collaboration} and {Amiri}, Mandana and {Andersen}, Bridget C. and {Andrew}, Shion and {Bandura}, Kevin and {Bhardwaj}, Mohit and {Boyle}, P.~J. and {Brar}, Charanjot and {Breitman}, Daniela and {Cassanelli}, Tomas and {Chawla}, Pragya and {Cook}, Amanda M. and {Curtin}, Alice P. and {Dobbs}, Matt and {Dong}, Fengqiu Adam and {Eadie}, Gwendolyn and {Fonseca}, Emmanuel and {Gaensler}, B.~M. and {Giri}, Utkarsh and {Herrera-Martin}, Antonio and {Hopkins}, Hans and {Ibik}, Adaeze L. and {Joseph}, Ronniy C. and {Kaczmarek}, J.~F. and {Kader}, Zarif and {Kaspi}, Victoria M. and {Lanman}, Adam E. and {Lazda}, Mattias and {Leung}, Calvin and {Liu}, Siqi and {Masui}, Kiyoshi W. and {McKinven}, Ryan and {Mena-Parra}, Juan and {Merryfield}, Marcus and {Michilli}, Daniele and {Ng}, Cherry and {Nimmo}, Kenzie and {Noble}, Gavin and {Pandhi}, Ayush and {Patel}, Chitrang and {Pearlman}, Aaron B. and {Pen}, Ue-Li and {Petroff}, Emily and {Pleunis}, Ziggy and {Rafiei-Ravandi}, Masoud and {Rahman}, Mubdi and {Ransom}, Scott M. and {Sand}, Ketan R. and {Scholz}, Paul and {Shah}, Vishwangi and {Shin}, Kaitlyn and {Shpunarska}, Yuliya and {Siegel}, Seth R. and {Smith}, Kendrick and {Stairs}, Ingrid and {Stenning}, David C. and {Vanderlinde}, Keith and {Wang}, Haochen and {White}, Henry and {Wulf}, Dallas},
        title = "{Updating the First CHIME/FRB Catalog of Fast Radio Bursts with Baseband Data}",
      journal = {\apj},
         year = 2024,
        month = jul,
       volume = {969},
       number = {2},
          eid = {145},
        pages = {145},
          doi = {10.3847/1538-4357/ad464b},
archivePrefix = {arXiv},
       eprint = {2311.00111},
 primaryClass = {astro-ph.HE},
       adsurl = {https://ui.adsabs.harvard.edu/abs/2024ApJ...969..145C}
}

@ARTICLE{Shannon2024,
       author = {{Shannon}, Ryan M. and {Bannister}, Keith W. and {Bera}, Apurba and {Bhandari}, Shivani and {Day}, Cherie K. and {Deller}, Adam T. and {Dial}, Tyson and {Dobie}, Dougal and {Ekers}, Ron D. and {Fong}, Wen-fai and {Glowacki}, Marcin and {Gordon}, Alexa C. and {Gourdji}, Kelly and {Jaini}, Akhil and {James}, Clancy W. and {Kumar}, Pravir and {Mahony}, Elizabeth K. and {Marnoch}, Lachlan and {Muller}, August R. and {Prochaska}, Xavier and {Qiu}, Hao and {Ryder}, Stuart D. and {Sadler}, Elaine M. and {Scott}, Danica R. and {Tejos}, N. and {Uttarkar}, Pavan A. and {Wang}, Yuanming},
        title = "{The commensal real-time ASKAP fast transient incoherent-sum survey}",
      journal = {\pasa},
         year = 2025,
        month = jan,
       volume = {42},
          eid = {e036},
        pages = {e036},
          doi = {10.1017/pasa.2025.8},
archivePrefix = {arXiv},
       eprint = {2408.02083},
 primaryClass = {astro-ph.HE},
       adsurl = {https://ui.adsabs.harvard.edu/abs/2025PASA...42...36S}
}

@ARTICLE{Connor2025,
       author = {{Connor}, Liam and {Ravi}, Vikram and {Sharma}, Kritti and {Ocker}, Stella Koch and {Faber}, Jakob and {Hallinan}, Gregg and {Harnach}, Charlie and {Hellbourg}, Greg and {Hobbs}, Rick and {Hodge}, David and {Hodges}, Mark and {Kosogorov}, Nikita and {Lamb}, James and {Law}, Casey and {Rasmussen}, Paul and {Sherman}, Myles and {Somalwar}, Jean and {Weinreb}, Sander and {Woody}, David and {Konietzka}, Ralf M.},
        title = "{A gas-rich cosmic web revealed by the partitioning of the missing baryons}",
      journal = {Nature Astronomy},
         year = 2025,
        month = aug,
       volume = {9},
        pages = {1226-1239},
          doi = {10.1038/s41550-025-02566-y},
archivePrefix = {arXiv},
       eprint = {2409.16952},
 primaryClass = {astro-ph.CO},
       adsurl = {https://ui.adsabs.harvard.edu/abs/2025NatAs...9.1226C}
}

@ARTICLE{Cui2025,
       author = {{Cui}, Xiang-han and {James}, Clancy. W. and {Li}, Di and {Zhang}, Cheng-min},
        title = "{Bias-corrected Fast Radio Burst Population and Spectra Using CHIME Injection Data}",
      journal = {\apj},
         year = 2025,
        month = apr,
       volume = {982},
       number = {2},
          eid = {158},
        pages = {158},
          doi = {10.3847/1538-4357/adbbcb},
archivePrefix = {arXiv},
       eprint = {2502.19138},
 primaryClass = {astro-ph.HE},
       adsurl = {https://ui.adsabs.harvard.edu/abs/2025ApJ...982..158C}
}

@ARTICLE{Pleunis2021morphology,
       author = {{Pleunis}, Ziggy and {Good}, Deborah C. and {Kaspi}, Victoria M. and {Mckinven}, Ryan and {Ransom}, Scott M. and {Scholz}, Paul and {Bandura}, Kevin and {Bhardwaj}, Mohit and {Boyle}, P.~J. and {Brar}, Charanjot and {Cassanelli}, Tomas and {Chawla}, Pragya and {(Adam) Dong}, Fengqiu and {Fonseca}, Emmanuel and {Gaensler}, B.~M. and {Josephy}, Alexander and {Kaczmarek}, Jane F. and {Leung}, Calvin and {Lin}, Hsiu-Hsien and {Masui}, Kiyoshi W. and {Mena-Parra}, Juan and {Michilli}, Daniele and {Ng}, Cherry and {Patel}, Chitrang and {Rafiei-Ravandi}, Masoud and {Rahman}, Mubdi and {Sanghavi}, Pranav and {Shin}, Kaitlyn and {Smith}, Kendrick M. and {Stairs}, Ingrid H. and {Tendulkar}, Shriharsh P.},
        title = "{Fast Radio Burst Morphology in the First CHIME/FRB Catalog}",
      journal = {\apj},
         year = 2021,
        month = dec,
       volume = {923},
       number = {1},
          eid = {1},
        pages = {1},
          doi = {10.3847/1538-4357/ac33ac},
archivePrefix = {arXiv},
       eprint = {2106.04356},
 primaryClass = {astro-ph.HE},
       adsurl = {https://ui.adsabs.harvard.edu/abs/2021ApJ...923....1P}
}

@ARTICLE{CHIME2023Cat1Reps,
       author = {{CHIME/Frb Collaboration} and {Andersen}, Bridget C. and {Bandura}, Kevin and {Bhardwaj}, Mohit and {Boyle}, P.~J. and {Brar}, Charanjot and {Cassanelli}, Tomas and {Chatterjee}, S. and {Chawla}, Pragya and {Cook}, Amanda M. and {Curtin}, Alice P. and {Dobbs}, Matt and {Dong}, Fengqiu Adam and {Faber}, Jakob T. and {Fandino}, Mateus and {Fonseca}, Emmanuel and {Gaensler}, B.~M. and {Giri}, Utkarsh and {Herrera-Martin}, Antonio and {Hill}, Alex S. and {Ibik}, Adaeze and {Josephy}, Alexander and {Kaczmarek}, Jane F. and {Kader}, Zarif and {Kaspi}, Victoria and {Landecker}, T.~L. and {Lanman}, Adam E. and {Lazda}, Mattias and {Leung}, Calvin and {Lin}, Hsiu-Hsien and {Masui}, Kiyoshi W. and {McKinven}, Ryan and {Mena-Parra}, Juan and {Meyers}, Bradley W. and {Michilli}, D. and {Ng}, Cherry and {Pandhi}, Ayush and {Pearlman}, Aaron B. and {Pen}, Ue-Li and {Petroff}, Emily and {Pleunis}, Ziggy and {Rafiei-Ravandi}, Masoud and {Rahman}, Mubdi and {Ransom}, Scott M. and {Renard}, Andre and {Sand}, Ketan R. and {Sanghavi}, Pranav and {Scholz}, Paul and {Shah}, Vishwangi and {Shin}, Kaitlyn and {Siegel}, Seth and {Smith}, Kendrick and {Stairs}, Ingrid and {Su}, Jianing and {Tendulkar}, Shriharsh P. and {Vanderlinde}, Keith and {Wang}, Haochen and {Wulf}, Dallas and {Zwaniga}, Andrew},
        title = "{CHIME/FRB Discovery of 25 Repeating Fast Radio Burst Sources}",
      journal = {\apj},
         year = 2023,
        month = apr,
       volume = {947},
       number = {2},
          eid = {83},
        pages = {83},
          doi = {10.3847/1538-4357/acc6c1},
archivePrefix = {arXiv},
       eprint = {2301.08762},
 primaryClass = {astro-ph.HE},
       adsurl = {https://ui.adsabs.harvard.edu/abs/2023ApJ...947...83C}
}

\appendix

\section{Modelling repeaters in ASKAP} \label{sec:askap_reps}

\begin{table}[]
    \centering
    \begin{tabular}{c|c c c}
   case &  $\log_{10} R_{\rm min}$ & $\log_{10} R_{\rm min}$ & $R_\gamma$ \\
       b  & -1.23 & -0.25 & -3 \\
       d  & -4.54 & 3 & -2.1 \\
    \end{tabular}
    \caption{Two cases of repetition parameters used for testing. Nomenclature from \citet{James2023}.}
    \label{tab:casebd}
\end{table}

The CRAFT survey with ASKAP is fully commensal, meaning that both the pointing position, and frequency configuration, change with time. This poses two challenges to modelling repeaters. In this section, we resolve these using two plausible scenarios (`case b' and `case d') for repeaters found by \citet{James2023}, with parameters reproduced in Table~\ref{tab:casebd}.

\subsection{Time per field}

Because the relative expected fraction of observed repeaters and single bursts is dependent upon observation time per field, the different dwell times per field for CRAFT formally requires a separately modelled survey for each and every pointing. Since this is computationally expensive, we here reduce the many different pointing times per field (\tfield) to a single effective pointing time (\teff) modelled identically over an effective number of fields (\neff). This then allows the total number of FRBs expected from our model to be linearly multiplied by \neff, since each field is independent.

\begin{figure}
    \centering
    \includegraphics[width=\linewidth]{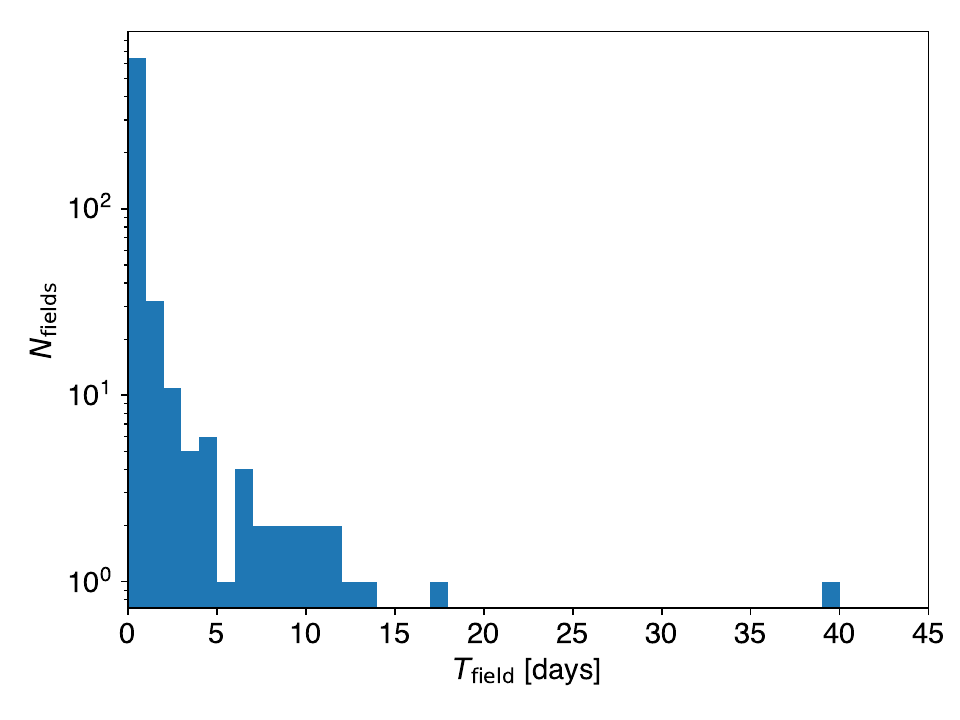}
    \caption{Histogram of the time spent observing each ASKAP field, proportional to the time spent observing each field.}
    \label{fig:tfield}
\end{figure}

\begin{figure}
    \centering
    \includegraphics[width=\linewidth]{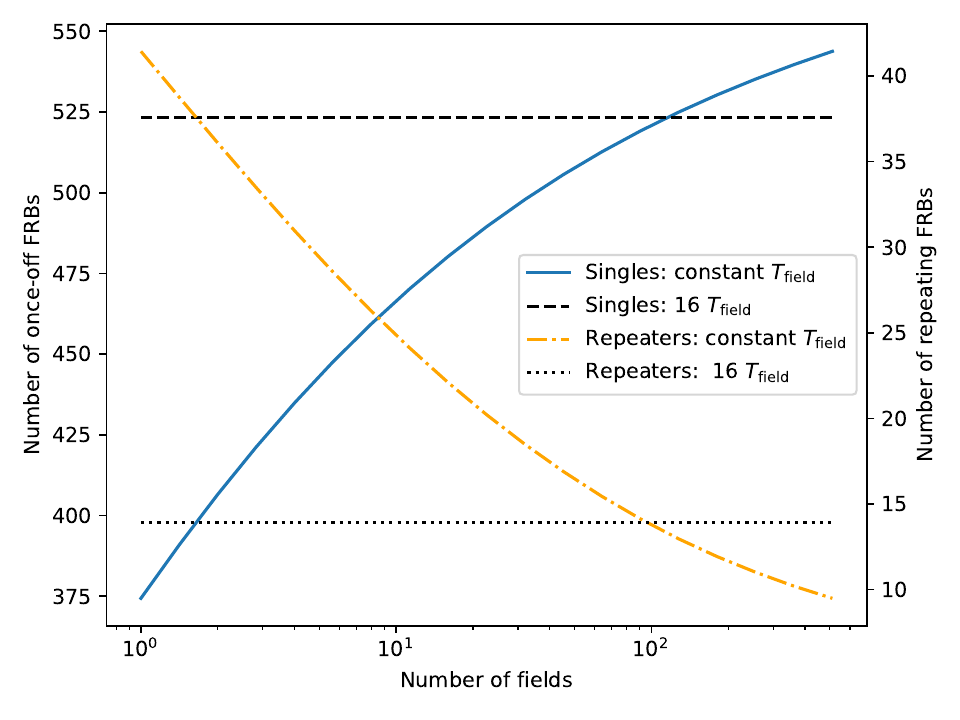}
    \caption{Number of single and repeat bursts as a function of the number of fields the ASKAP ICS survey is assumed to have covered at constant time \tfield, compared to the more-detailed estimates using 16 values of \tfield\ according to Figure~\ref{fig:tfield}. This model assumes repetition parameters from case b (Table \ref{tab:casebd}) and best-fit FRB population parameters from \citet{Hoffmann2025}. }
    \label{fig:fit_tfield}
\end{figure}

A histogram of the number of fields, as a function of pointing time per field, is shown in Figure~\ref{fig:tfield}.
Grouping the fields by \tfield\ in one-day increments as per that figure, we sum the expected number of repeaters in cases b and d over all 16 values of \tfield. We then model a single value of \tfield, and vary the number of fields from 1 to 512, setting \tfield\ to keep the total survey time constant. The resulting number of single and repeat FRBs are plotting in Figure~\ref{fig:fit_tfield}. We find for cases b and d respectively that a single effective pointing time of 4.3 and 3.2\,days --- corresponding to 105 and 141 pointings --- reproduces the correct ratio of single and repeat FRBs. We therefore choose to describe the ASKAP observations as 123 pointings, each with an observation time of 3.66 days.

\subsection{Frequency configuration}

The different frequency configurations of the CRAFT survey poses a less tractable issue. In previous works, we have treated these frequency configurations by grouping them into low ($< 1$\,GHz), mid ($1-1.4$\,GHz), and high ($> 1.4$\,GHz) configurations. However, this is not feasible when dealing with repeaters, since the \zdm\ code necessarily treats surveys independently. Yet surveys will be mutually dependent in the case that a field is searched at multiple different frequencies, since the cosmic variance of repeater density will be a common factor between them. For example --- if a strong repeater is not in a field when searched at mid frequencies, there will be no strong repeater in that field when searches at low or high frequencies. Indeed, this co-dependence between surveys is also a factor between ASKAP, Parkes, and CRAFT, which all have partially overlapping fields of view. However, the effect will be much weaker between different instruments, which have concentrated their searches on different parts of the sky.

A full treatment of this effect could only be done using Monte Carlo methods --- we see no full analytic or semi-analytic treatment within a code such as \zdm{} as being possible. Rather, we here make an approximation that allows us to treat all ASKAP configurations as being performed at one effective frequency.

In \citet{Shannon2024}, a toy model was developed that modelled the FRB rate in terms of changing ASKAP frequency and beam configuration. The model assumed a Euclidean fluence distribution $dN/dF \propto F^{-2.5}$, which allows the rate dependence of the varying number of antennas used in the incoherent sum, and different beam configurations, to be accounted-for; a dependence of detection rate on frequency as $\nu^{-\alpha}$, where we now use $\alpha=-1.5$; and the sensitivity effects due to the effective width of FRBs changing with frequency and sampling time, where we use the measured width and scattering times of FRBs from \citep{Shannon2024} to estimate sensitivity-dependent effects. Here, we normalise the weightings produced by this model to ASKAP's closepack36 beam configuration with $0.9^{\circ}$ beam spacing at 1.272\,GHz and integration time of 1.182\,ms. This produces a total effective observation time of 522.5 days.

\begin{figure}
    \centering
    \includegraphics[width=\linewidth]{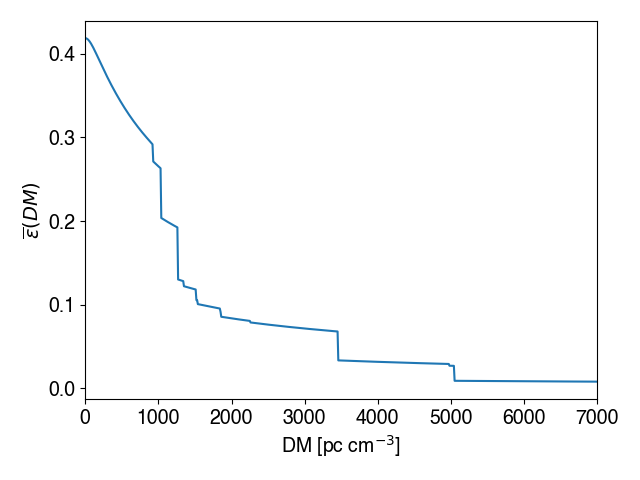}
    \caption{Efficiency value of ASKAP FRB observations as a function of DM, averaged over all observing configurations. The sudden jumps are due to ASKAP's maximum search DM varying as a function of frequency.}
    \label{fig:dm_response}
\end{figure}

We also use a DM-dependent response, which is a weighted sum over all DM response curves generated by ASKAP surveys. This is plotted in Figure~\ref{fig:dm_response}, and is implemented in the simulation using the methods of \citet{Hoffmann2024}.

\begin{figure}
    \centering
    \includegraphics[width=\linewidth]{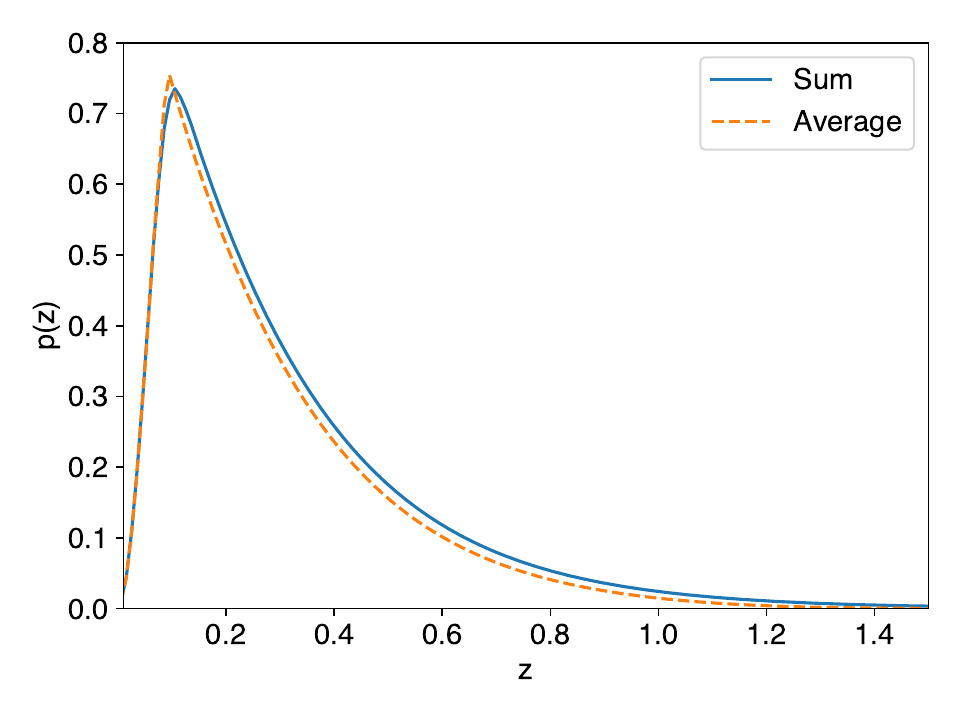}
    \caption{Comparison of the redshift distribution of FRBs generated using the averaged ASKAP observations with that from the sum of low-, medium-, and high-band observations as per \citet{james2022}.}
    \label{fig:z_comparison}
\end{figure}

To check that our averaged value of ASKAP sufficiently reproduces the properties of the previously used ASKAP surveys, which were grouped into low ($<1$\,GHz), medium ($1 < < 1.4$\,GHz , and high ($>1.4$\,GHz) bands, we compare the predicted redshift distribution of single bursts assuming no repetition in Figure~\ref{fig:z_comparison}. The average distribution is slightly left-shifted compared to that summed over the three frequency ranges; however, this difference is smaller than the difference we find when iterating over different FRB parameter sets. We therefore conclude that our approximation is reasonable.

\section{Results including $P(N)$} \label{sec:Pn}
\begin{figure*}
\begin{center}
\includegraphics[width=\linewidth]{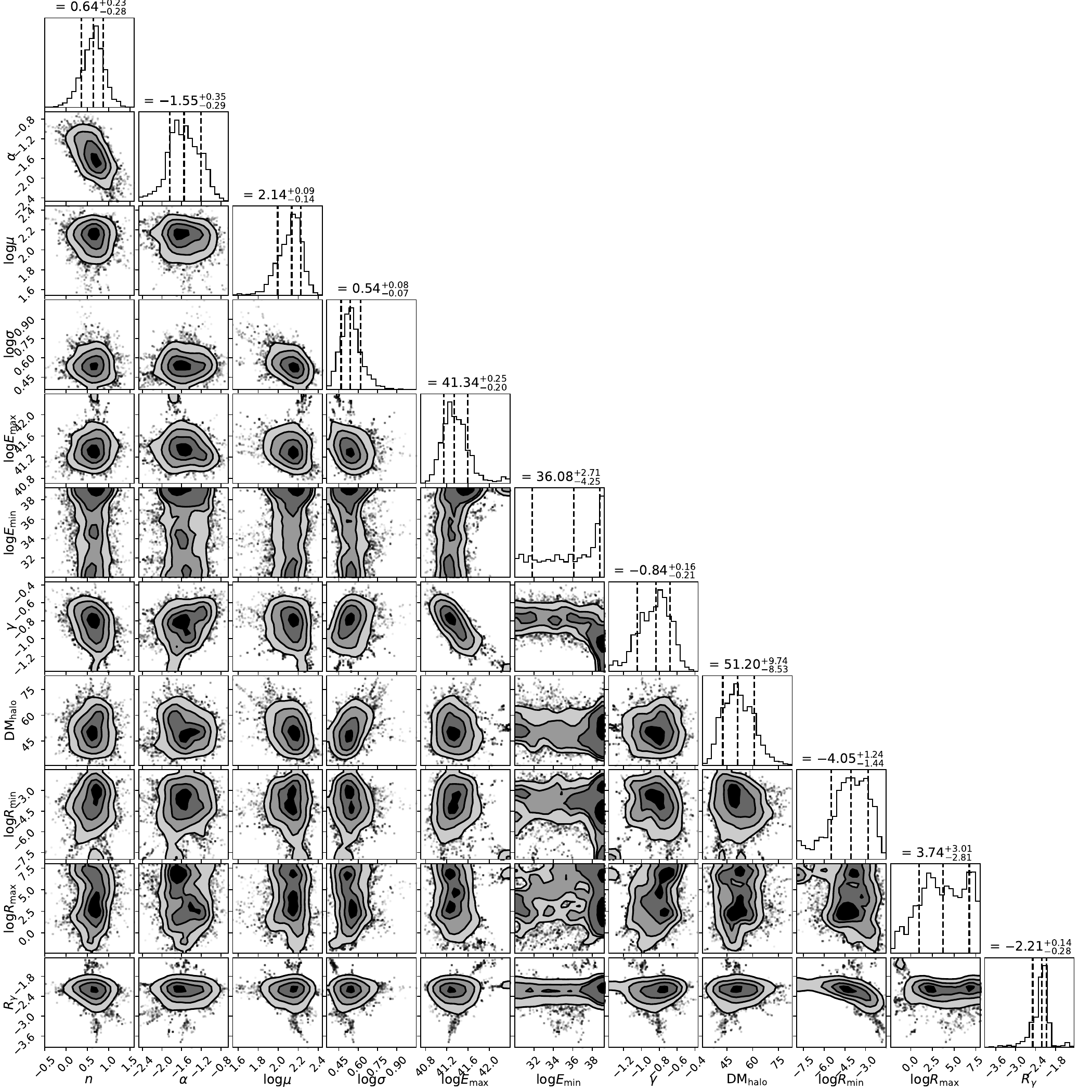}
\caption[]{Results from the MCMC analysis, including repetition parameters. These results use the same method as those shown in the main results of Figure \ref{fig:cornerplot}, but also include $P(N)$ for Murriyang (Parkes), \flyseye{} and FAST.}
\label{fig:cornerplotPN}
\end{center}
\vspace{-3ex}
\end{figure*}

Due to the limitations we have in determining accurate observation times for each survey and inconsistencies even from the same instrument, we have chosen to exclude $P(N)$ in the main results of this paper. In this section, we present our results when we do include $P(N)$ for the reader's information. These results are shown in Figure \ref{fig:cornerplotPN}. This figure is identical to the main results shown in Figure \ref{fig:cornerplot} but additionally includes $P(N)$ for the non-repetition surveys.

Most parameters do not show significant differences with the inclusion of $P(N)$. The only noteworthy change is in $\alpha$, which is in line with expectations. Constraints on the spectral behaviour rely primarily on the relative abundance of FRBs at different frequencies. As each telescope operates within a limited frequency band, the relative abundance of FRBs detected by different surveys becomes a key factor. However, to determine the relative abundance at different frequencies, the individual detection biases of each telescope and total observation time must be disentangled. Thus, not having reliable estimates for this means there are large uncertainties in the number of expected FRBs within a given survey, which will have a significant impact on the value of $\alpha$ we obtain. However, this error is not reflected in our quoted uncertainties.

\end{document}